\documentclass[aps,prd]{revtex4-2}
\usepackage{cancel}
\usepackage[normalem]{ulem}
\usepackage{graphicx}
\usepackage{float}
\usepackage{amsmath}
\usepackage{slashed}
\usepackage{xcolor}
\usepackage{tabularx}
\usepackage{multirow}
\usepackage{caption}
\usepackage{subfigure}
\usepackage{morefloats}

\makeatletter
\newcommand*{\rom}[1]{\expandafter\@slowromancap\romannumeral #1@}
\makeatother

\begin{document}
\title{Coupled-channel description of charmed heavy hadronic molecules within the meson-exchange model and its implication}
\author{Lin Qiu$^{1,2}$\footnote{{\it E-mail address:} qiulin@ihep.ac.cn},
Chang Gong$^{1,4}$\footnote{{\it E-mail address:} gongchang@ccnu.edu.cn},
Qiang Zhao$^{1,2,3}$\footnote{{\it E-mail address:} zhaoq@ihep.ac.cn}}
\affiliation{$^1$ Institute of High Energy Physics and Theoretical Physics Center
for Science Facilities,\\
         Chinese Academy of Sciences, Beijing 100049, China}

\affiliation{$^2$ School of Physical Sciences, University of Chinese Academy of
Sciences, Beijing 100049, China}

\affiliation{$^3$ China Center of Advanced Science and Technology, Chinese Academy of
Sciences, Beijing 100080, China}

\affiliation{$^4$ Key Laboratory of Quark and Lepton Physics (MOE) and Institute of Particle Physics, Central China Normal University, Wuhan 430079, China}

\date{\today}
\begin{abstract}
    Motivated by the first observation of the double-charm tetraquark $T_{cc}^+(3875)$ by the LHCb Collaboration, we investigate the nature of $T_{cc}^+$ as an isoscalar $DD^*$ hadronic molecule in a meson-exchange potential model incorporated by the coupled-channel effects and three-body unitarity. The $D^0D^0\pi^+$ invariant mass spectrum can be well-described and the $T_{cc}^+$ pole structure can be precisely extracted. Under the hypothesis that the interactions between the heavy flavor hadrons can be saturated by the light meson-exchange potentials, the near-threshold dynamics of $T_{cc}^+$ can shed light on the binding of its heavy-quark spin symmetry (HQSS) partner $D^*D^*$ ($I=0$) and on the nature of other heavy hadronic molecule candidates such as $X(3872)$ and $Z_c(3900)$ in the charmed-anticharmed systems. The latter states can be related to $T_{cc}^+$ in the meson-exchange potential model with limited assumptions based on the SU(3) flavor symmetry relations. The combined analysis, on the one hand, indicates the HQSS breaking effects among those HQSS partners, and on the other hand, highlights the role played by the short and long-distance dynamics for the near threshold $D^{(*)}D^{(*)}$ and $D^{(*)}\bar{D}^{(*)}+c.c.$ systems.
\end{abstract}

\maketitle

\section{Introduction}
One of the critical issues about the non-perturbative property of quantum chromodynamics (QCD) is to what extent it allows the existence of the so-called ``exotic hadrons" of which the constituent contents are beyond the conventional quark model (i.e. mesons are composed of $q\bar{q}$ and baryons of $qqq$~\cite{Gell-Mann:1964ewy,Zweig:1964ruk,Zweig:1964jf,Godfrey:1985xj,Capstick:1986ter}). Such exotic hadrons include glueballs, hybrids, multiquarks, and hadronic molecules, etc, and their existences may serve as a unique probe for understanding the non-perturbative property of QCD. Since the discovery of $X(3872)$ by the Belle Collaboration in 2003~\cite{Belle:2003nnu}, there have been a large number of exotic candidates observed in experiments (see e.g. Refs.~\cite{Guo:2017jvc,Chen:2016qju,Olsen:2017bmm,Esposito:2016noz,Lebed:2016hpi,Brambilla:2019esw,Liu:2019zoy,Ali:2017jda,Karliner:2017qhf,Chen:2022asf} for recent reviews). Interestingly, most of these observed states are heavy flavor states and located in the vicinity of some relative $S$-wave thresholds. It makes the hadronic molecule picture a natural solution for their nature, and also allows the implementation of effective field theory (EFT) approaches in the description of the near-threshold dynamics. Such a phenomenon is very similar to that for the deuteron system, where the EFT approach originally proposed by Weinberg~\cite{Weinberg:1965zz} have been greatly developed during the past decades~\cite{Weinberg:1990rz,Weinberg:1991um,Kaplan:1998tg,Kaplan:1998we}.

From the perspective of the EFTs, the non-perturbative property of QCD can be learned from different aspects at different low energy scales. In the light-quark sector, the light meson-meson interactions can be described by the chiral perturbation theory (ChPT), which is constructed by the pseudo-Nambu-Goldstone boson fields that emerge from the spontaneous symmetry breaking of chiral symmetry. While in the heavy-heavy sector, the interactions between two heavy hadrons are well constrained by the heavy quark symmetry (HQS) and heavy quark spin symmetry (HQSS). Due to the large heavy quark masses ($m_Q\gg \Lambda_{\text{QCD}}$) the mid and short-range interactions can be described by only several low energy constants (LECs) for different isospin channels. However, these LECs still cannot be well-determined due to the lack of experimental data and taking into account also the uncertainty $\mathcal{O}(\Lambda_{\text{QCD}}/m_Q)$ if $m_Q$ is not large enough. 

From another view, the hadron-hadron interactions can also be described by the potentials generated by the meson exchanges, such as scalar ($\sigma, \ \kappa$), vector ($\rho, \ \omega, \ J/\psi$), pseudoscalar mesons ($\pi, \ \eta, \ \eta_c$), etc. They are similar to the meson exchange potentials between nucleons inside a nucleus and can mimic the underlying quark-exchange processes. With the decay constants determined by the on-shell heavy-meson decays, these LECs are hypothesized to be saturated by matching the ChPT operators expansion with those meson-exchange potentials~\cite{Ecker:1988te,Donoghue:1988ed}.

With the interactions mentioned above, the dynamical mechanism between hadron-hadron pairs can then be investigated and the complete scattering amplitudes can be accessed by making a proper analytical continuation of the on-shell ones into the complex plane with the exception of finite singularities determined by the particle-exchange kinematics~\cite{mandelstam:1958,PhysRev.126.1596}. Then the resonance peaks observed in the invariant mass spectrum may be derived from the poles of the scattering amplitudes on the complex plane, which has been proved successful in the description of many low-energy meson-meson scatterings~\cite{GomezNicola:2007qj,GomezNicola:2001as} and meson (nucleon)-nucleon scatterings~\cite{Schutz:1998jx,Krehl:1999km} in the light-quark sector. As to the very long-range potential mediated by the light pion exchange, it brings into the breaking of HQS/HQSS, and due to the on-shell decay $D^*\rightarrow D\pi$ the two-hadron composite systems are usually quasi-bound/virtual-states-like resonances. Strictly, the multi-hadrons involved processes can be learned by solving the Faddeev-type equations~\cite{faddeev:1965}. But in practice they are often reduced to an effective two-body Lippmann-Schwinger equation with the assumption that the two-body interaction proceeds via an isobar~\cite{Aaron:1968aoz,Janssen:1994uf}. In such a way, the three-body unitarity is reserved and the poles of these hadronic molecules acquire a certain imaginary part which arises from these sub-threshold cuts.

In 2021, the LHCb Collaboration announced the first observation of a double-charm tetraquark $T_{cc}^+(3875)$ ($cc\bar{u}\bar{d}$) in the $D^0D^0\pi^+$ invariant mass distribution~\cite{LHCb:2021vvq} and its Breit-Wigner parameters are,
\begin{equation*}
    \delta m_{\text{BW}}=-273\pm 61 \ \text{keV}/c^2, \ \Gamma_{\text{BW}}=410\pm 165 \ \text{keV} ,
\end{equation*}
with $\delta m_{\text{BW}}$ the mass relative to the nominal $D^0D^{*+}$ threshold and $\Gamma_{\text{BW}}$ the width. Later, a unitarized analysis by LHCb Collaboration~\cite{LHCb:2021auc} suggests it is a pole on the second Riemann sheet,
\begin{equation*}
    \delta m_{\text{pole}}=-360\pm 40^{+4}_{-0} \ \text{keV}/c^2, \ \Gamma_{\text{pole}}=48\pm 2^{+0}_{-14} \ \text{keV},
\end{equation*}
where $\delta m_{\text{pole}}$ also refers to the real part of its pole relative to the $D^0D^{*+}$ threshold and $\Gamma_{\text{pole}}$ is twice of the absolute value of  the imaginary part of its pole. The partial amplitude analysis and the absence of its isospin partners in $D^+D^0\pi^+$ channel imply its quantum numbers as $IJ^{P}=01^+$. Besides, the fact that approximately $90\%$ of the $D^0D^0\pi^+$ events contain a genuine $D^{*+}$ meson reveals its main components as $D^0D^{*+}$, and it should couple strongly to $D^+D^{*0}$ and $DD\pi$ due to the proximity among those channel thresholds. Early theoretical studies of the double heavy-flavor exotic states can be found in the literature~\cite{Zouzou:1986qh,Lipkin:1986dw,Heller:1986bt,Carlson:1987hh,Silvestre-Brac:1993zem,Pepin:1996id,Brink:1998as,Janc:2004qn,Zhang:2007mu,Barnea:2006sd,Vijande:2007fc,Navarra:2007yw,Manohar:1992nd,Carames:2011zz,Ohkoda:2012hv}.

With the discovery of $T_{cc}^+(3875)$, a lot of efforts have been done on understanding its nature based on different scenarios, such as compact states~\cite{Kim:2022mpa,Chen:2022ros}, hadronic molecules~\cite{Feijoo:2021ppq,Dong:2021bvy,Dai:2021wxi,Albaladejo:2021vln,Du:2021zzh,Ke:2021rxd,Abreu:2022sra,Xin:2021wcr,Peng:2023lfw}, triangle singularity (TS) mechanism~\cite{Braaten:2022elw}, etc. It was also investigated by Lattice QCD (LQCD) calculations~\cite{Padmanath:2022cvl,Chen:2022vpo,Lyu:2023xro,Du:2023hlu}, and in its production and decays~\cite{Hu:2021gdg,Ling:2021bir,Meng:2021jnw,Yan:2021wdl}. More or less, these models can provide a reasonable explanation of its formation and production. The existence of $X(3872)$, $T_{cc}^+$ and $Z_c(3900)$ suggests peculiar dynamics arising from the $D^{(*)}\bar{D}^{(*)}$ and $D^{(*)}D^{(*)}$ threshold interactions. Thus, a combined analysis of these states, which involves the $D^{(*)}\bar{D}^{(*)}$ and $D^{(*)}D^{(*)}$ threshold interactions, is necessary for gaining deeper insights into the underlying dynamics.

Different from the double-charm system, the charmed-anticharmed systems have already been studied broadly in both experiment and theory. Although the nature of these resonances is still heatedly debated due to the more complicated analytical structures. In our work, we first study the near-threshold dynamical nature of $T_{cc}^+$ as an isoscalar $DD^*$ hadronic molecule within the meson-exchange model including both coupled-channel effects and the three-body unitarity. Then, we try to explore other possibilities of hadronic molecules composed by charm-anticharm heavy mesons such as $X(3872)$ and $Z_c(3900)$ which can be regarded as partners of $T_{cc}^+$ with respect to flavour symmetry and charge conjugate transformation beyond the HQS/HQSS in a more general basis. In this sense, our combined analysis of these systems tends to, on the one hand, demonstrate the relations based on the HQS/HQSS, and on the other hand, manifest the effects caused by the HQS/HQSS breaking in the $D^{(*)}\bar{D}^{(*)}$ and $D^{(*)}D^{(*)}$ interactions. 

This paper is organized as follows: In Sec.~\ref{sec:fra}, we discuss the potentials constructed by the one-boson-exchange (OBE) model and coupled-channel formalism, and the eigen wavefunctions satisfying the Hamiltonian. In Sec.~\ref{sec:res}, we present our numerical results and discussions. A brief summary is given in Sec.~\ref{sec:con}.

\section{Framework}
\label{sec:fra}

\subsection{One-boson-exchange model and potentials}
The OBE model describes the hadron interactions similar to that for the nuclear force. The exchanged bosons, which have different quantum numbers and masses, account for a lot of features of the nuclear elements. While the boson exchanges are actually the quark exchanges between hadrons, the OBE model can be regarded as a leading order approximation of the strong QCD in the non-perturbative regime. 
Within the OBE model the dynamics for the interacting hadrons are generally fine-tuned by a physical cutoff $\Lambda$ ($0.3\sim 1.2$ GeV) by matching the theoretical calculations to the experimental data. The cutoff indicates a typical scale beyond which the boson exchange scenario does not hold anymore. The  relevance of these dynamical ingredients can be learned by the so-called Weinberg organizational principle~\cite{Weinberg:1990rz,Weinberg:1991um,Kaplan:1998tg,Kaplan:1998we} which efficiently collects the soft/hard scales and counts the operators order by order. 

In the heavy meson-(anti) meson sector~\cite{Valderrama:2012jv}, it can be learned that the one-pion-exchange (OPE) and coupled-channel effects (i.e. the $DD-DD^*-D^*D^*$ couplings for the double-charm system), are both next-to-leading-order (NLO). In contrast, the leading-order (LO) contributions are from the light vector meson exchanges, which are treated as dynamical gauge bosons of hidden local symmetries~\cite{Bando:1984ej,Bando:1987br,Bando:1985rf,Meissner:1987ge,Nagahiro:2008cv}.  At this moment, we limit our analysis in the sector of non-strange charmed meson systems. We note that in the bottom sector the large masses of $B^{(*)}$ mesons can bring in a large typical relative momentum for the $B^{(*)}B$ ($B^{(*)}\bar{B}$) systems at the mass region of the $B^* B^*$ ($B^*\bar{B}^*$) threshold though the threshold differences among $B^{(*)}B^{(*)}$ or $(B^{(*)}\bar{B}^{(*)})$ channels are rather small, i.e. $\delta< m_\pi$ with $\delta\equiv M_{B^*}-M_B\simeq 50$ MeV. As the result, the OPE contribution is no longer perturbative and can be promoted to the LO~\cite{Wang:2018jlv}.  This will slow down the convergence of the EFT. For composite systems with strangeness, the SU(3) flavour symmetry becomes the approximate one and the kaon-induced interactions are very subtle due to lightness of kaon mesons.

The LO Lagrangians for the S-wave heavy (anti-)mesons interactions respecting the heavy quark symmetry and the SU(3)-flavor symmetry read~\cite{Wise:1992hn,Falk:1992cx,Grinstein:1992qt,Casalbuoni:1996pg},
\begin{align}
\label{eq:lagobe}
    \begin{split}
    \mathcal{L}_{HH\Pi}&=ig\langle H_b^{(Q)}\gamma_{\mu}\gamma_5A_{ba}^{\mu}\bar{H}_a^{(Q)}\rangle+ig\langle \bar{H}_a^{(\bar{Q})}\gamma_{\mu}\gamma_5 A_{ab}^{\mu}H_b^{(\bar{Q})}\rangle+\cdots\\
    \mathcal{L}_{HHV}&=i\beta\langle H_b^{(Q)}\nu_{\mu}(V_{ba}^{\mu}-\rho_{ba}^{\mu})\bar{H}_a^{(Q)}\rangle+i\lambda \langle H_b^{(Q)}\sigma_{\mu\nu}F^{\mu\nu}(\rho)_{ba}\bar{H}_a^{(Q)}\rangle\\
    &-i\beta\langle \bar{H}_a^{(\bar{Q})}\nu_{\mu}(V_{ab}^{\mu}-\rho_{ab}^{\mu})H_b^{(\bar{Q})}\rangle+i\lambda\langle \bar{H}_a^{(\bar{Q})}\sigma_{\mu\nu}F^{\mu\nu}H_b^{(\bar{Q})}\rangle+\cdots\\
    \mathcal{L}_{HH\sigma}&=g_s\langle H_a^{(Q)}\sigma \bar{H}_a^{(Q)}\rangle +g_s\langle \bar{H}_a^{(\bar{Q})}\sigma H_a^{(\bar{Q})}\rangle+\cdots \ ,
    \end{split}
\end{align}
where $\langle\cdots\rangle$ denotes tracing over the Dirac $\gamma$ matrices, $H_a^{(Q)}$ and $\bar{H}_{a}^{(Q)}$ are the superfields that annihilate and create heavy mesons, respectively. Similarly, $H_a^{(\bar{Q})}$ and $\bar{H}_a^{(\bar{Q})}$ are the superfields that annihilate and create heavy anti-mesons, respectively. They have the following expressions:
\begin{align}
    H_a^{(Q)}&=\frac{1+\slashed{\nu}}{2}[P_a^{*(Q)\mu}\gamma_{\mu}-P_a^{(Q)}\gamma_5], \\
    \bar{H}_a^{(Q)}&=\gamma_0 H_a^{(Q)\dagger}\gamma_0=[P_a^{*(Q)\dagger\mu}\gamma_{\mu}+P_a^{(Q)\dagger}\gamma_5]\frac{1+\slashed{\nu}}{2}, \\
    H_{a}^{(\bar{Q})}&=C(\mathcal{C}H_a^{(Q)}\mathcal{C}^{-1})^{T}C^{-1}=[P_a^{*(\bar{Q})\mu}\gamma_{\mu}-P_a^{(\bar{Q})}\gamma_5]\frac{1-\slashed{\nu}}{2}, \\ \bar{H}_a^{(\bar{Q})}&=\gamma_0 H_{a}^{(\bar{Q})\dagger}\gamma_0=\frac{1-\slashed{\nu}}{2}[P_a^{*(\bar{Q})\mu}\gamma_{\mu}+P_a^{(\bar{Q})}\gamma_5] \ ,
\end{align}
with $\mathcal{C}$ the charge conjugation operator and $C=i\gamma_2\gamma_0$ the charge conjugation matrix. In the charmed meson sector $P^{(Q)}=(D^0,D^+,D_s^+)$ and $P^{*(Q)}=(D^{*0},D^{*+},D_s^{*+})$ are the pseudoscalar and vector charmed mesons along with their anti-charmed ones $P^{(\bar{Q})}=(\bar{D}^0,D^-,D_s^-)$ and $P^{*(\bar{Q})}=(\bar{D}^{*0},D^{*-},D_s^{*-})$. The axial-current $A^{\mu}$ is defined as $A^{\mu}=\frac{1}{2}(\xi^{\dagger}\partial^{\mu}\xi -\xi\partial^{\mu}\xi^{\dagger})$, where $\xi=e^{\frac{i\Pi}{f_{\pi}}}$ with $\Pi$ the pseudoscalar meson fields given in Eq.~(\ref{eq:current}). The vector current is $V^{\mu}=\frac{1}{2}(\xi^{\dagger}\partial^{\mu}\xi+\xi\partial^{\mu}\xi^{\dagger})$. In the heavy quark limit $\nu^{\mu}=p^{\mu}/M=(1,0,0,0)$ is the four-velocity of the heavy meson and $F^{\mu\nu}(\rho)=\partial^{\mu}\rho^{\nu}-\partial^{\nu}\rho^{\mu}-[\rho^{\mu},\rho^{\nu}]$ with $\rho^{\mu}$ the vector meson fields given in the following equation:
\begin{align}
    \label{eq:current}
    \begin{split}
        \Pi&=\begin{pmatrix}
        \frac{1}{\sqrt{2}}\pi^{0}+\frac{\eta}{\sqrt{6}}&\pi^{+}&K^{+}\\
        \pi^{-}&-\frac{1}{\sqrt{2}}\pi^{0}+\frac{\eta}{\sqrt{6}}&K^{0}\\
        K^{-}&\bar{K}^{0}&-\frac{2}{\sqrt{6}}\eta
        \end{pmatrix} \ ,\\
        \rho&=i\frac{g_{V}}{\sqrt{2}}\hat{\rho}=i\frac{g_{V}}{\sqrt{2}}\begin{pmatrix}
        \frac{\rho^{0}}{\sqrt{2}}+\frac{\omega}{\sqrt{2}}&\rho^{+}&K^{*+}\\
        \rho^{-}&-\frac{\rho^{0}}{\sqrt{2}}+\frac{\omega}{\sqrt{2}}&K^{*0}\\
        K^{*-}&\bar{K}^{*0}&\phi
        \end{pmatrix} .
    \end{split}
\end{align}
The Lagrangians in Eq.~(\ref{eq:lagobe}) can thus be expanded as,
\begin{align}
    \begin{split}
    \mathcal{L}&= -i\frac{2g}{f_{\pi}}\epsilon_{\alpha\mu\nu\lambda}\nu^{\alpha}P^{*(Q)\mu}_{b}P^{*(Q)\lambda\dagger}_{a}\partial^{\nu}\Pi_{ba}-\frac{2g}{f_{\pi}}(P_b^{(Q)}P^{*(Q)\dagger}_{a\lambda}+P^{*(Q)}_{b\lambda}P^{(Q)\dagger}_{a})\partial^{\lambda}\Pi_{ba}-\sqrt{2}\beta g_VP_b^{(Q)}P_a^{(Q)\dagger}\nu\cdot \hat{\rho}_{ba}\\
    &-2\sqrt{2}\lambda g_{V}\nu^{\lambda}\epsilon_{\lambda\mu\alpha\beta}(P_{b}^{(Q)}P_{a}^{*(Q)\mu\dagger}+P_{b}^{*(Q)\mu}P_{a}^{(Q)\dagger})(\partial^{\alpha}\hat{\rho}^{\beta})_{ba}+\sqrt{2}\beta g_{V}P_{b}^{*(Q)}\cdot P_{a}^{*(Q)\dagger}\nu\cdot \hat{\rho}_{ba}\\
    &-i2\sqrt{2}\lambda g_{V}P_{b}^{*(Q)\mu}P_{a}^{*(Q)\nu\dagger}(\partial_{\mu}\hat{\rho}_{\nu}-\partial_{\nu}\hat{\rho}_{\mu})_{ba}-2g_{s}P_{b}^{(Q)}P_{b}^{(Q)\dagger}\sigma+2g_{s}P_{b}^{*(Q)}\cdot P_{b}^{*(Q)\dagger}\sigma\\
    &+i\frac{2g}{f_{\pi}}\epsilon_{\alpha\mu\nu\lambda}\nu^{\alpha}P^{*(\bar{Q})\mu\dagger}_{a}P^{*(\bar{Q})\lambda}_{b}\partial^{\nu}\Pi_{ab}+\frac{2g}{f_{\pi}}(P_{a\lambda}^{*(\bar{Q})\dagger}P^{(\bar{Q})}_{b}+P^{(\bar{Q})\dagger}_{a}P^{*(\bar{Q})}_{b\lambda})\partial^{\lambda}\Pi_{ab}+\sqrt{2}\beta g_VP_a^{(\bar{Q})\dagger}P_b^{(\bar{Q})}\nu\cdot \hat{\rho}_{ab}\\
    &-2\sqrt{2}\lambda g_{V}\nu^{\lambda}\epsilon_{\lambda\mu\alpha\beta}(P_{a}^{*(\bar{Q})\mu\dagger}P_{b}^{(\bar{Q})}+P_{a}^{(\bar{Q})\dagger}P_{b}^{*(\bar{Q})\mu})(\partial^{\alpha}\hat{\rho}^{\beta})_{ab}-\sqrt{2}\beta g_{V}P_{a}^{*(\bar{Q})\dagger}\cdot P_{b}^{*(\bar{Q})}\nu\cdot \hat{\rho}_{ab}\\
    &-i2\sqrt{2}\lambda g_{V}P_{a}^{*(\bar{Q})\mu\dagger}P_{b}^{*(\bar{Q})\nu}(\partial_{\mu}\hat{\rho}_{\nu}-\partial_{\nu}\hat{\rho}_{\mu})_{ab}-2g_{s}P_{a}^{(\bar{Q})\dagger}P_{a}^{(\bar{Q})}\sigma+2g_{s}P_{a}^{(\bar{Q})*\dagger}\cdot P_{a}^{*(\bar{Q})}\sigma+\cdots 
    \end{split}
\end{align}
In this work we adopt the same coupling constants as used in Ref.~\cite{PhysRevD.88.114008}, i.e. the pion decay constant $f_{\pi}=132$ MeV, $g=0.59\pm 0.007\pm 0.01$, $g_{V}=5.8$, $\beta=0.9$, $\lambda=0.56\text{GeV}^{-1}$ and $g_{s}=g_{\pi}/(2\sqrt{6})$ with $g_{\pi}=3.73$. For Lagrangians involving the exchange of heavy mesons ($J/\psi$, $\eta_{c}$, etc.), they can be derived by replacing the SU(3) flavour octets by the ones in the SU(4) flavour symmetry~\cite{Hofmann:2005sw,Ikeno:2020mra}.

For composite system $DD^{*}$ in an $S$-wave, the wave functions with quantum numbers $J^P=1^{+}$ with $I=0, \ 1$ can be constructed as,
\begin{align}
    \begin{split}
   &I=0: \frac{1}{\sqrt{2}}(\bar{u}\bar{d}-\bar{d}\bar{u})\otimes\frac{1}{\sqrt{2}}(PV+VP)=\frac{1}{\sqrt{2}}(\frac{1}{\sqrt{2}}(|D^0D^{*+}\rangle-|D^{*+}D^0\rangle)-\frac{1}{\sqrt{2}}(|D^+D^{*0}\rangle-|D^{*0}D^+\rangle)) \\
   &I=1: \frac{1}{\sqrt{2}}(\bar{u}\bar{d}+\bar{d}\bar{u})\otimes\frac{1}{\sqrt{2}}(PV+VP)=\frac{1}{\sqrt{2}}(\frac{1}{\sqrt{2}}(|D^0D^{*+}\rangle+|D^{*+}D^0\rangle)+\frac{1}{\sqrt{2}}(|D^+D^{*0}\rangle+|D^{*0}D^+\rangle)) \ .
   \end{split}
\end{align}
It should be noted that the systems of $D^{0}D^{*0} \ (I=1, \ I_3=-1)$ and $D^{+}D^{*+} \ (I=1, \ I_3=+1)$ are just partners of the $DD^{*}$ isospin triplet with $(I=1, \ I_3=0)$. In view of the  hidden local symmetries the isoscalar system is found to be attractive, and the isovector one turns out to be repulsive with the light vector meson exchanges. Thus, we consider only the above wave functions for $DD^*$ and charge-neutral ones for the charmed-anticharmed system.

Note that the full eigenstate must include all the degrees of freedom. By solving the two-body problem, one obtains the eigenstate for this two-body system, but would not be able to distinguish $D^*$ and $D$. The eigenstate describes the relative motions of the two-body system quantum-mechanically in a relative spatial separation of $r$ (corresponding to the momentum transfer between the two constituents in the momentum space) without knowing which one is which. Similar considerations were also discussed in Ref.~\cite{Thomas:2008ja}. In our recent work~\cite{gong}, we have made a detailed survey of the wave functions and potentials of the heavy hadronic molecules in accordance with their quantum numbers and we briefly summarize the basic ingredients here for reference. This issue is also noticed by Ref.~\cite{Ke:2021rxd}.

Moreover, we emphasize that the total wave function, which is symmetric under space, isospin and spin group $O(3)\otimes SU_I(2)\otimes SU_S(2)$ for bosons, is crucial for properly introducing the meson-exchange potentials, especially the pion-exchange ones from the $u$ channel. Since the full pionful interaction requires the inclusion of loop corrections~\cite{Wang:2013kva,Baru:2015nea} and the pion-exchange potential would not account for the short-distance dynamics with $r<<1/m_\pi$, different parametrizations of the short-distance potential can result in different behaviors of the nonperturbative OPE to be either attractive or repulsive~\cite{Valderrama:2012jv}. This also calls for a unified approaches for the $D^{(*)}\bar{D}^{(*)}$ and $D^{(*)}D^{(*)}$ systems. Otherwise, the conclusion about the importance of OPE could become self-contradicting and even misleading. 

For the $S$-wave $D\bar{D}^{*}$ composite system, the charge-neutral wave functions for different isospins and charge conjugations can be constructed in a similar way,
\begin{align}
    \begin{split}
   &I=0, \ C=+: \frac{1}{\sqrt{2}}(\bar{u}u+\bar{d}d)\otimes\frac{1}{\sqrt{2}}(P\bar{V}-V\bar{P})=\frac{1}{\sqrt{2}}(\frac{1}{\sqrt{2}}(D^0\bar{D}^{*0}-D^{*0}\bar{D}^0)+\frac{1}{\sqrt{2}}(D^+D^{*-}-D^{*+}D^-)) , \\
   &I=0, \ C=-: \frac{1}{\sqrt{2}}(\bar{u}u+\bar{d}d)\otimes\frac{1}{\sqrt{2}}(P\bar{V}+V\bar{P})=\frac{1}{\sqrt{2}}(\frac{1}{\sqrt{2}}(D^0\bar{D}^{*0}+D^{*0}\bar{D}^0)+\frac{1}{\sqrt{2}}(D^+D^{*-}+D^{*+}D^-)) , \\
   &I=1, \ C=+: \frac{1}{\sqrt{2}}(\bar{u}u-\bar{d}d)\otimes\frac{1}{\sqrt{2}}(P\bar{V}-V\bar{P})=\frac{1}{\sqrt{2}}(\frac{1}{\sqrt{2}}(D^0\bar{D}^{*0}-D^{*0}\bar{D}^0)-\frac{1}{\sqrt{2}}(D^+D^{*-}-D^{*+}D^-)) , \\
   &I=1, \ C=-: \frac{1}{\sqrt{2}}(\bar{u}u-\bar{d}d)\otimes\frac{1}{\sqrt{2}}(P\bar{V}+V\bar{P})=\frac{1}{\sqrt{2}}(\frac{1}{\sqrt{2}}(D^0\bar{D}^{*0}+D^{*0}\bar{D}^0)-\frac{1}{\sqrt{2}}(D^+D^{*-}+D^{*+}D^-)) .
   \end{split}
\end{align}

Since for many of those hadronic molecules, the binding energy is far smaller than the threshold difference between coupled channels and thus the isospin breaking effect should be taken into account. Therefore, we define two channels in the particle basis as in Ref.~\cite{PhysRevD.88.114008}:  $[D^0D^{*+}]_{\mp}  =(|D^0D^{*+}\rangle\mp |D^{*+}D^0\rangle)/{\sqrt{2}}$ and $[D^+D^{*0}]_{\mp}  =(|D^+D^{*0}\rangle\mp |D^{*0}D^+\rangle)/{\sqrt{2}}$. The physical state $T_{cc}^{+}$ is a linear combination of $([D^{0}D^{*+}]_{-}-[D^{+}D^{*0}]_{-})/\sqrt{2}$ (rather than $([D^{0}D^{*+}]_{+}-[D^{+}D^{*0}]_{+})/\sqrt{2}$ used in Refs.~\cite{PhysRevD.88.114008,Chen:2021vhg,Cheng:2022qcm}) taking into account the coupled-channel effects. Similarly, we define two channels for $D\bar{D}^{*}$: $[D^0\bar{D}^{*0}]_{\mp}=(D^0\bar{D}^{*0}\mp D^{*0}\bar{D}^0)/{\sqrt{2}}$ and $[D^+D^{*-}]_{\mp}=(D^{+}D^{*-}\mp D^{*+}D^{-})/{\sqrt{2}}$ among which $X(3872)$ is described by $[D\bar{D}^{*}]_{-}$ and $Z_c(3900)$ by $[D\bar{D}^{*}]_{+}$.

The near-threshold potentials can then be evaluated by the static approximation~\cite{Ericson:1988gk}: 
\begin{equation}\hat{V}(ab\rightarrow cd)=-\frac{\hat{\mathcal{M}}(ab\rightarrow cd)}{\sqrt{\prod_i(2m_i)\prod_f(2m_f)}} \ ,
\end{equation}
with $\hat{\mathcal{M}}$ the scattering amplitude of the process $ab\rightarrow cd$ and $m_{i/f}$ the masses of initial or final state mesons. For convenience, we first define the following functions:
\begin{align}
    \begin{split}
        \tilde{\mathcal{X}}_{\text{ex}}&=\frac{\boldsymbol{\epsilon}_i\cdot\boldsymbol{\epsilon}^*_f}{|\boldsymbol{q}|^2+m_{\text{ex}}^2-{q^{0}}^2}\\
        \mathcal{X}_{\text{ex}}&=\frac{(1-\frac{{q^0}^2}{m_{\text{ex}}^2})\boldsymbol{\epsilon}_i\cdot\boldsymbol{\epsilon}^*_f}{|\boldsymbol{q}|^2+m_{\text{ex}}^2-{q^{0}}^2}\\
        \mathcal{Y}_{\text{ex}}&=\frac{\boldsymbol{\epsilon}_i\cdot\boldsymbol{q}\boldsymbol{\epsilon}^*_f\cdot\boldsymbol{q}}{|\boldsymbol{q}|^2+m_{\text{ex}}^2-{q^{0}}^2}\\
        \mathcal{Z}_{\text{ex}}&=\frac{(\boldsymbol{\epsilon}_i\times\boldsymbol{q})\cdot(\boldsymbol{\epsilon}^*_f\times\boldsymbol{q})}{|\boldsymbol{q}|^2+m_{\text{ex}}^2-{q^{0}}^2}
    \end{split}
\end{align}
where $\boldsymbol{\epsilon}_i$ is the initial polarization vector of $D^*(\bar{D}^*)$, $\boldsymbol{\epsilon}^*_f$ is the final one, $m_{\text{ex}}$ is the mass of the exchanged meson, and $q^0$/$\boldsymbol{q}$ are the zero-th/three-vector components of the transferred momentum $q^\mu=p^{\prime\mu}-p^\mu$ with $p^{\prime\mu}/p^\mu$ the momenta of the final/initial  mesons. Interestingly, $q^0$ could be larger than the mass of the exchanged pion for some processes like $D^{0}D^{*+}\xrightarrow{\pi^-}D^{*+}D^0$. It will lead to a logarithmic divergence in the projected partial waves and should be properly treated by a full consideration of the three-body unitarity~\cite{Baru:2011rs,PhysRevD.105.014024}.

\begin{figure}[H]
    \centering
    \includegraphics[width=0.4\textwidth]{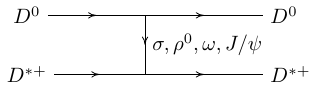}
    \includegraphics[width=0.4\textwidth]{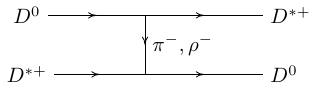}
    \caption{Direct (left) and crossing (right) feynmann diagrams of process $D^0D^{*+}\rightarrow D^{0}D^{*+}$.} 
    \label{fig:feydig}
\end{figure}

The detailed potentials for $DD^*\rightarrow DD^*$ (as depicted in Fig.\ref{fig:feydig}) are listed as follows:
\begin{enumerate}
    \item $D^0D^{*+}\rightarrow D^0D^{*+}$/$D^+D^{*0}\rightarrow D^+D^{*0}$
    \begin{align}
        \begin{split}
            V^{[DD^*]_{\mp}}&=-\frac{g_V^2\beta^2}{4}\mathcal{X}_{\rho^0}+\frac{g_V^2\beta^2}{4}\mathcal{X}_{\omega}-g_S^2\tilde{\mathcal{X}}_{\sigma}\pm \frac{g^2}{f_{\pi}^2}\mathcal{Y}_{\pi^{-}}\pm 2g_V^2\lambda^2\mathcal{Z}_{\rho^{-}}+\frac{g_V^2\beta^2}{2}\mathcal{X}_{J/\psi}
        \end{split}
    \end{align}
    \item $D^0D^{*+}\rightarrow D^+D^{*0}$
    \begin{align}
        \begin{split}
            V^{[DD^*]_{\mp}}&=\frac{g_V^2\beta^2}{2}\mathcal{X}_{\rho^-}\mp\frac{g^2}{2f_{\pi}^2}\mathcal{Y}_{\pi^0}\pm\frac{g^2}{6f_{\pi}^2}\mathcal{Y}_{\eta}\mp g_V^2\lambda^2\mathcal{Z}_{\rho^0}\pm g_V^2\lambda^2\mathcal{Z}_{\omega}
            \pm\frac{g^2}{f_{\pi}^2}\mathcal{Y}_{\eta_c}\pm 2g_V^2\lambda^2\mathcal{Z}_{J/\psi}
        \end{split}
    \end{align}
\end{enumerate}
Similarly for $D\bar{D}^*$,
\begin{enumerate}
    \item $D^0\bar{D}^{*0}\rightarrow D^0\bar{D}^{*0}$/$D^+D^{*-}\rightarrow D^+D^{*-}$
    \begin{align}
        \begin{split}
            V^{[D\bar{D}^*]_{\mp}}&=-\frac{g_V^2\beta^2}{4}\mathcal{X}_{\rho^0}-\frac{g_V^2\beta^2}{4}\mathcal{X}_{\omega}-g_S^2\tilde{\mathcal{X}}_{\sigma}\mp\frac{g^2}{2f_{\pi}^2}\mathcal{Y}_{\pi^0}\mp\frac{g^2}{6f_{\pi}^2}\mathcal{Y}_{\eta}\pm\lambda^2g_V^2\mathcal{Z}_{\rho^0}\pm\lambda^2g_V^2\mathcal{Z}_{\omega}\\
            &\mp \frac{g^2}{f_{\pi}^2}\mathcal{Y}_{\eta_c}-\frac{g_V^2\beta^2}{2}\mathcal{X}_{J/\psi}\pm 2\lambda^2g_V^2\mathcal{Z}_{J/\psi}
        \end{split}
    \end{align}
    \item $D^0\bar{D}^{*0}\rightarrow D^+D^{*-}$
    \begin{align}
        \begin{split}
            V^{[D\bar{D}^{*}]_{\mp}}&=-\frac{g_V^2\beta^2}{2}\mathcal{X}_{\rho^{-}}\mp\frac{g^2}{f_{\pi}^2}\mathcal{Y}_{\pi^{-}}\pm 2g_V^2\lambda^2\mathcal{Z}_{\rho^{-}}
        \end{split}
    \end{align}
\end{enumerate}

To saturate the LECs with the meson-exchange potentials, one matches the contact-range operator expansion with those potentials generated by the light meson exchange~\cite{Peng:2020xrf}. We illustrate this pattern with OPE by decomposing it into two terms,
\begin{align}
    \mathcal{Y}_{\pi}&=\frac{1}{3}\{(1-\frac{\mu_{\pi}^2}{|\boldsymbol{q}|^2+\mu_{\pi}^2})\hat{S}(\boldsymbol{\epsilon}_i,\boldsymbol{\epsilon}_f^*)+\frac{|\boldsymbol{q}|^2}{|\boldsymbol{q}|^2+\mu_{\pi}^2}\hat{T}(\boldsymbol{\epsilon}_i,\boldsymbol{\epsilon}_f^*)\}
\end{align}
where $\mu_{\pi}^2=m_{\pi}^2-{q^0}^2$, the first term is called central term $\hat{S}(\boldsymbol{\epsilon}_i,\boldsymbol{\epsilon}_f^*)=\boldsymbol{\epsilon}_i\cdot\boldsymbol{\epsilon_f^*}$ and the second term is called tensor term $\hat{T}(\boldsymbol{\epsilon}_i,\boldsymbol{\epsilon}_f^*)=3\boldsymbol{\epsilon}_i\cdot\boldsymbol{\hat{q}}\boldsymbol{\epsilon}^*_f\cdot\boldsymbol{\hat{q}}-\boldsymbol{\epsilon}_i\cdot\boldsymbol{\epsilon}_f^*|\boldsymbol{\hat{q}}|^2$ with $\boldsymbol{\hat{q}}$ the unit of $\boldsymbol{q}$. Because of the perturbative contribution by the pion-exchange diagrams, the tensor term is negligible and we only consider the $S$-wave component in our work. For the central term, the first Dirac delta term actually is unphysical since the dynamics of $\delta(r)$ is far beyond the reach of the pion scale $1/m_{\pi}$. It is generally parametrized by a finite size $\sim 1/\Lambda$ with $\Lambda >m_\pi$. In our approach, we parametrize this term along with the heavier pseudoscalar ($P$) and vector ($V$) meson ($m_{P,V}\sim\Lambda$) exchange potentials via the scale-dependent counter terms $C_{P,V}$, which are not fully captured by the OBE model~\cite{Yalikun:2021bfm}. Thus, we adopt the following substitution rules~\cite{Ji:2022blw},
\begin{align}
    \mathcal{Y}_{P/V}&\sim \frac{1}{3}(-\frac{\mu_{P/V}^2}{|\boldsymbol{q}|^2+\mu_{P/V}^2}+C_{P/V})\hat{S}(\boldsymbol{\epsilon}_i,\boldsymbol{\epsilon}_f^*) \ ,\\
    \mathcal{Z}_{V}&\sim \frac{2}{3}(-\frac{\mu_V^2}{|\boldsymbol{q}|^2+\mu_V^2}+C_V)\hat{S}(\boldsymbol{\epsilon}_i,\boldsymbol{\epsilon}_f^*)
\end{align}
with $\mu_i^2=m_i^2-{q^0}^2$ and $0\leq C_{P/V}\leq 1$. For very heavy  exchanged mesons (e.g. $J/\psi, \eta_c$), we just ignore the spin-dependent terms $\mathcal{Y}_{\text{ex}},\mathcal{Z}_{\text{ex}}$ as they vanish at the threshold (since the Dirac delta terms are fully reserved), but keep the resumed Yukawa terms $\mathcal{X}_{\text{ex}}$. We collect the wave functions and potentials of other heavy double-charm or charmed-anticharmed systems in Appendix~\ref{appendix.a}.

\subsection{Lippmann-Schwinger equation and three-body interactions}

The near-threshold coupled-channel dynamics can be studied by solving the nonrelativistical Lippmann-Schwinger equation (LSE),
\begin{align}
    \label{eq:lse}
    \begin{split}
        T_{\alpha\beta}(\boldsymbol{p},\boldsymbol{k};E)&=V_{\alpha\beta}(\boldsymbol{p},\boldsymbol{k};E)+\sum_{\delta}\int\frac{d^3\boldsymbol{q}}{(2\pi)^3}V_{\alpha\delta}(\boldsymbol{p},\boldsymbol{q};E)G_{\delta\delta}(\boldsymbol{q};E)T_{\delta\beta}(\boldsymbol{q},\boldsymbol{k};E)
    \end{split}    
\end{align}
where $E$ is the energy in the initial c.m. frame, $V_{\alpha\beta}(\boldsymbol{p},\boldsymbol{k};E)$ is the potential from the $\beta$-th channel to the $\alpha$-th channel. To regularize the ultra-violet (UV) divergence we introduce a hard cutoff into the LSE by replacing the potentials with,
\begin{equation}
    V_{\alpha\beta}(\boldsymbol{p},\boldsymbol{q};E)\rightarrow V_{\alpha\beta}(\boldsymbol{p},\boldsymbol{q};E)\Theta(\Lambda-|\boldsymbol{p}|)\Theta(\Lambda-|\boldsymbol{q}|)
\end{equation}
with $\Theta$ the Heaviside step function and $\Lambda$ the cutoff parameter. We refrain from the use of any Gaussian or monopole form factors because of their complexities in the undermentioned analytical continuation~\cite{Sadasivan:2021emk}. $G_{\delta\delta}(\boldsymbol{q};E)$ is the two-body propagator of the $\delta$-th channel which can be written as (in the nonrelativistical limit),
\begin{align}
    G_{\delta\delta}(\boldsymbol{q};E)=\frac{1}{E-m_1-m_2-\frac{\boldsymbol{q}^2}{2\mu_{\delta}}+i\frac{\Gamma_{\delta}(\boldsymbol{q};E)}{2}}
\end{align}
with $m_i$ the mass of the constituent particle, $\mu_{\delta}$ the reduced mass of the $\delta$-th channel and $\Gamma_{\delta}(\boldsymbol{q};E)$ the width contribution of the $\delta$-th channel. As we only focus on the near-threshold phenomena (e.g. $3.873\sim 3.877$ GeV for the $D^0D^0\pi^+$ mass spectrum), the above nonrelativistic approximation should be reasonable. Since the width of $D^*$ is comparable with the width of the corresponding hadronic molecules, namely $T_{cc}^+$ and $X(3872)$, the width part of the propagator $\Gamma_{\delta}$ will have a significant effect on the pole position. In most of the literature, the constant width treatment for $D^*$ might be appropriate as argued in Ref.~\cite{Kang:2016jxw}. But the width of the composite state could be overestimated by two times, compared with the one in which the full three-body unitarity is considered~\cite{Baru:2011rs}. In the case of a constant width for $D^*$, the Riemann sheets can be classified according to the sign of the imaginary parts of the on-shell momenta (taking two channels $D^{0}D^{*+}$ and $D^{+}D^{*0}$ as example): 
\begin{equation}
    k_1=\sqrt{2\mu_{1}(E-m_{D^0}-m_{D^{*+}}+i\frac{\Gamma_{D^{*+}}}{2})},\quad k_2=\sqrt{2\mu_{2}(E-m_{D^+}-m_{D^{*0}}+i\frac{\Gamma_{D^{*0}}}{2})}
\end{equation}
with $\mu_1$, $\mu_2$ the reduced masses of the $D^{0}D^{*+}$, $D^{+}D^{*0}$ channels, respectively. Namely,
\begin{align}
    \begin{split}
        &{\text{RS-\rom{1}}}: \Im k_1>0, \Im k_2>0 \ , \\
        &{\text{RS-\rom{2}}}: \Im k_1<0, \Im k_2>0 \ , \\
        &{\text{RS-\rom{3}}}: \Im k_1<0, \Im k_2<0 \ , \\
        &{\text{RS-\rom{4}}}: \Im k_1>0, \Im k_2<0.
    \end{split}\label{eq:twors}
\end{align}

\begin{figure}[H]
  \subfigure[]{
        \includegraphics[width=5cm]{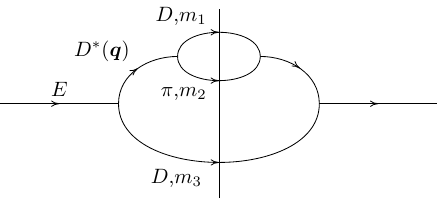}\label{fig:three-body1}}
%
  \subfigure[]{
        \includegraphics[width=12cm]{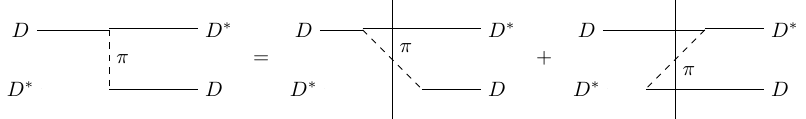}\label{fig:three-body2}}
%
    \caption{Three-body cuts involved in the $DD^*$ scattering. (a) Self energy of $D^*$ by the $D\pi$ loop; (b) On-shell decay of process $D^*\rightarrow D\pi$. The left hand side is the propagator of pion in Feynmann representation used in our work and the right hand side is the sum of the forward and backward emissions of pion in TOPT.}
    \label{fig:three-body}
\end{figure}

Apart from the LO tree diagrams at the order $\mathcal{Q}^{-1}$, the NLO diagrams involving the pion exchange which can contribute to the decay width of composite particle $DD^*(\bar{D}^*)$, is depicted in Fig.~\ref{fig:three-body}. They are suppressed by one power of $\mathcal{Q}$ as analyzed with an effective Lagrangian derived from heavy-hadron chiral perturbation theory (HH$\chi$PT)~\cite{Fleming:2007rp}. While the NLO contribution is also found to be dominant by the contact potential, the wave function receives the renormalization, especially the imaginary part, by evaluating the above cut diagrams which are characterized by the $D\pi$ loop and one-pion-exchange. Regardless of the full-pion calculation, the nonperturbative nature of the OPE was justified by the opposing argument about the calculation of one-loop diagrams using the dimensional regularization scheme~\cite{Wang:2013kva} and the sharp cutoff scheme~\cite{Baru:2015nea} which leads to the conclusion that the OPE potential in an EFT is only well-defined in connection with a contact term. Moreover, the width of $X(3872)$ with the effects of three-body cuts and its quark-mass-dependence were also validated in the EFT treatment with both perturbative pions~\cite{Fleming:2007rp,Jansen:2013cba} and nonperturbative pions~\cite{Baru:2011rs}. In summary, the $DD^*$ scattering in three-body unitarity must include the full $D^*$ propagators and the pion-exchange diagrams in the on-shell renormalization scheme. 

In our approach since the dominant contact part in the OPE is completely removed, the perturbative nature of OPE is retained if the physical cutoff and other counter terms are fine-tuned to the full line shape. We leave the discussion about its analytical treatment in Appendix~\ref{appendix.b}. On the other hand, the width of $D^*$ should be energy-dependent due to the self-energy from $D\pi$ loop as done in Refs.~\cite{Baru:2011rs,PhysRevD.105.014024,Feijoo:2021ppq,Ji:2022blw}. For $D^*\rightarrow D\pi$, the width contribution to the complex mass of $D^*$ is,
\begin{equation}
    \Gamma\propto p_{cm}^3 \ ,
\end{equation}
where $p_{cm}=\sqrt{2\mu_{D\pi}(E'-m_D-m_{\pi})}$ is the magnitude of three-momentum in the c.m. frame of $D^*$ with $E'$ the energy in the c.m. frame of $D^*$ and $m_D$ and $m_{\pi}$ the masses of $D$ and pion, respectively, as depicted in Fig.~\ref{fig:three-body1}. The first/second Riemann sheets are also defined regarding the imaginary part of $p_{cm}$ induced by the right-hand cut (RHC) lying along [$m_D+m_{\pi}$,$\infty$), i.e.,
\begin{align}
    \begin{split}
        &\text{RS-\rom{1}}: \Im p_{cm}>0 \ ,\\
        &\text{RS-\rom{2}}: \Im p_{cm}<0 .
    \end{split}
\end{align}

In the moving frame of $D^*(\boldsymbol{q})$, the energy available for $D^*$ gets reduced since the denominator of the $D\pi$ propagator is now,
\begin{align}
    \begin{split}
        &E'-\sqrt{m_D^2+(\boldsymbol{q}/2+\boldsymbol{l})^2}-\sqrt{m_{\pi}^2+(\boldsymbol{q}/2-\boldsymbol{l})^2}\\
        &=E'-\omega_{D}-\omega_{\pi}-(1/\omega_{D}+1/\omega_{\pi})\frac{\boldsymbol{q}^2}{8}+\mathcal{O}(\boldsymbol{q}^4) \ ,
    \end{split}
\end{align}
where $\omega_i=\sqrt{m_i^2+\boldsymbol{l}^2}$ is the energy of the constituent particle in the static c.m. frame of $D^*$($\boldsymbol{q}=\boldsymbol{0}$) with $\boldsymbol{l}$ the relative three-momentum between $D$ and $\pi$ in the $D\pi$ loop. In the framework of Time-Ordered-Perturbation-Theory (TOPT), the static energy of $D^*$ in the initial c.m. frame can be approximated by
\begin{equation}
    \label{eq:boost}
    E'(E,\boldsymbol{q})=E+m_{D^*}-\sqrt{m_{D^*}^2+\boldsymbol{q}^2}-\sqrt{m_{D}^2+\boldsymbol{q}^2}
\end{equation}
with $m_{D^*}$ the bare mass of $D^*$.

We denote the $p_{cm}$ solved from the above process by: $p_{cm}=\mathcal{F}(E,\boldsymbol{q},m_1,m_2,m_3)$ and then the energy-dependent width of $D^{*+}$/$D^{*0}$ is (analogous to Refs.~\cite{Baru:2011rs,PhysRevD.105.014024}),
\begin{align}
    \label{eq:dswidth}
    \begin{split}
        \Gamma_c(\boldsymbol{q};E)&=\Gamma(D^{*+}\rightarrow D^{+}\gamma)+\frac{g^2m_{D^0}}{6\pi f_{\pi}^2m_{D^{*+}}}\mathcal{F}^3(E,\boldsymbol{q},m_{D^{0}},m_{\pi^+},m_{D^0})+\frac{g^2m_{D^+}}{12\pi f_{\pi}^2m_{D^{*+}}}\mathcal{F}^3(E,\boldsymbol{q},m_{D^+},m_{\pi^0},m_{D^0}) \ ,\\
        \Gamma_0(\boldsymbol{q};E)&=\Gamma(D^{*0}\rightarrow D^0\gamma)+\frac{g^2m_{D^0}}{12\pi f_{\pi}^2 m_{D^{*0}}}\mathcal{F}^3(E,\boldsymbol{q},m_{D^0},m_{\pi^0},m_{D^+})+\frac{g^2m_{D^+}}{6\pi f_{\pi}^2m_{D^{*0}}}[\mathcal{F}^3(E,\boldsymbol{q},m_{D^+},m_{\pi^-},m_{D^+})\\
        &-\mathcal{F}^3(m_{D^+}+m_{D^{*0}},\boldsymbol{0},m_{D^+},m_{\pi^-},m_{D^+})].
    \end{split}
\end{align}

For the purpose of searching for poles, the function $\mathcal{F}(E,\boldsymbol{q},m_1,m_2,m_3)$ has to be analytically and properly continued by the technique of redirecting the branch cuts and contour deformation. We only present the result in this section and refer to Refs.~\cite{Doring:2009yv,PhysRevC.84.015205,PhysRevD.105.014024,Sadasivan:2021emk} or Appendix~\ref{appendix.b} for details,
\begin{equation}
    \label{eq:self-energy-ac}
    \tilde{\mathcal{F}}(E,\boldsymbol{q},m_1,m_2,m_3)=\begin{cases}
        -\mathcal{F}(E,\boldsymbol{q},m_1,m_2,m_3),\quad\text{if }\Im(E'(E,\boldsymbol{q}))<0\text{ and }\Re(E'(E,\boldsymbol{q}))>m_1+m_2 \ ,\\
        \mathcal{F}(E,\boldsymbol{q},m_1,m_2,m_3),\quad\text{else}.
    \end{cases}
\end{equation}

\subsection{Line shape analysis of the $D^0D^0\pi^+$ mass spectrum}
\begin{figure}[H]
    \centering
    \includegraphics{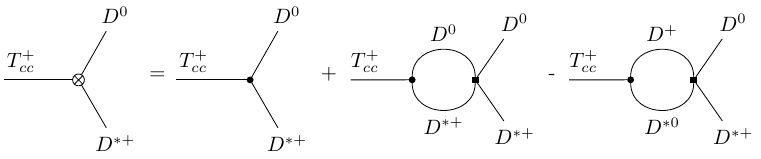}
    \caption{Feynman diagram for the $T_{cc}^+\rightarrow D^0D^{*+}$ transitions.} 
    \label{fig:prod}
\end{figure}

We solve the Lippmann-Schwinger equation in the particle basis,
\begin{equation}
    \{D^0D^{*+}(^3S_1),D^+D^{*0}(^3S_1)\} \ ,
\end{equation}
and the LO Lagrangian to describe the interaction between $T_{cc}^+$ and $D/D^*$ is,
\begin{equation}
    \mathcal{L}_{XDD^*}=ig_{XDD^*}(X^{\mu}D^{*+}_{\mu}D^{0}\mp X^{\mu}D^{*0}_{\mu}D^{+}) \ ,
\end{equation}
where symbol $\mp$ corresponds to the isoscalar/isovector $DD^*$ system, and $g_{XDD^{*}}$ is the coupling constant and approximated by a constant due to the narrow $D^0D^0\pi^+$ invariant mass range $\sim 4$ MeV. Note that the contribution from the $D$ wave is neglected due to the suppression of the centrifugal barrier. Thus, the transition $T_{cc}^+\rightarrow D^0D^{*+}$ can be evaluated by (as depicted by Fig.~\ref{fig:prod}),
\begin{align}
    t(\boldsymbol{p};E)&=g_{XDD^*}+g_{XDD^*}\int\frac{d^3\boldsymbol{q}}{(2\pi)^3}T_{11}(\boldsymbol{p},\boldsymbol{q};E)G_{11}(\boldsymbol{q};E)\\
    &-g_{XDD^*}\int\frac{d^3\boldsymbol{q}}{(2\pi)^3}T_{12}(\boldsymbol{p},\boldsymbol{q};E)G_{22}(\boldsymbol{q};E)
\end{align}
with $\boldsymbol{p}$ the three momentum of $D^0$ in the c.m. frame of $T_{cc}^+$. The coupling constant $g_{XDD^{*}}$ can be absorbed into the overall factor $\mathcal{N}$ and thus is set to unit. 

\begin{figure}[H]
    \centering
    \includegraphics{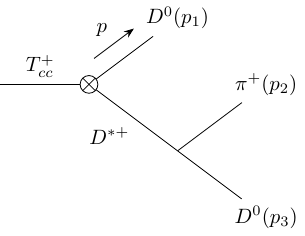}
    \caption{Tree-level diagram for $T_{cc}^+\rightarrow D^0\pi^+D^0$.}
    \label{fig:tree}
\end{figure}
The decay width of $T_{cc}^+\rightarrow D^0D^0\pi^+$ is (see Fig.~\ref{fig:tree}),
\begin{align}
    \begin{split}
        \frac{d\Gamma_{T_{cc}^+\rightarrow D^0D^0\pi^+}(E)}{ds_{12}ds_{23}}\propto |q_{\pi}G_{11}(p;E)t(p;E)+\bar{q}_{\pi}G_{11}(\bar{p};E)t(\bar{p};E)|^2 \ ,
    \end{split}
\end{align}
where the two terms come from the symmetry of two $D^0$ mesons in the final state with $q_{\pi} \ (\bar{q}_{\pi})$ the magnitude of three-momentum of pion in the c.m. frame of $D^{*+}[\pi^+(p_2)D^0(p_3)]$ and $p \ (\bar{p})$ the magnitude of the $D^0$ three-momentum produced at the first vertex in the initial c.m. frame, namely,
\begin{align}
    \begin{split}
        q_{\pi}&=\frac{\sqrt{\lambda(s_{23},m_{\pi^+}^2,m_{D^0}^2)}}{2\sqrt{s_{23}}},\quad p=\frac{\sqrt{\lambda(E^2,m_{D^0}^2,s_{23})}}{2E} \ ,\\
        \bar{q}_{\pi}&=\frac{\sqrt{\lambda(s_{12},m_{D^0}^2,m_{\pi^{+}}^2)}}{2\sqrt{s_{12}}},\quad \bar{p}=\frac{\sqrt{\lambda(E^2,m_{D^0}^2,s_{12})}}{2E} \ ,
    \end{split}
\end{align}
where $\lambda(x,y,z)=x^2+y^2+z^2-2xy-2xz-2yz$ is the Källén function, $s_{12}=(p_1+p_2)^2$, $s_{23}=(p_2+p_3)^2$ are the invariant masses squared, and the integral limits are determined by the Dalitz boundary,
\begin{align}
    \begin{split}
        (m_1+m_2)^2&\leq s_{12}\leq (E-m_3)^2 \ ,\\
        (E_{2}^*+E_3^*)^2-(\sqrt{E_2^*-m_2^2}+\sqrt{E_3^2-m_3^2})^2&\leq s_{23}\leq (E_{2}^*+E_3^*)^2-(\sqrt{E_2^*-m_2^2}-\sqrt{E_3^2-m_3^2})^2
    \end{split}
\end{align}
with $m_1$, $m_2$, and $m_3$ the masses of $D^0$, $\pi^+$, and $D^0$, respectively, and $E_2^*\equiv \frac{s_{12}-m_1^2+m_2^2}{2\sqrt{s_{12}}}$ and $E_3^*\equiv \frac{E^2-s_{12}-m_3^2}{2\sqrt{s_{12}}}$.

In order to compare with the experimental data, the decay width function above should be convoluted with the mass resolution function of the LHCb detector, which are modulated by the sum of two Gaussian functions~\cite{Dai:2021wxi},
\begin{align}
    \label{eq:yields}
    \begin{split}
         \frac{\text{Yields}}{\Delta E}&=\mathcal{N}\int_{E_i-\Delta E/2}^{E_i+\Delta E/2}dE \frac{\Gamma_{T_{cc}^+\rightarrow D^0D^0\pi^+}(E)}{\Delta E}\{\sum_{j=1}^2\beta_j\cdot \frac{1}{\sqrt{2\pi}\sigma_j}\text{exp}[-\frac{1}{2}(\frac{E-E_i}{\sigma_j})^2]\} \ ,
    \end{split}
\end{align}
where $\Delta E=200$ keV is the bin width; $\beta_1=0.778$,  $\beta_2=0.222$, $\sigma_1=1.05\times 263$ keV, and $\sigma_2=2.413\times\sigma_1$ are taken from the LHCb analysis~\cite{LHCb:2021vvq,LHCb:2021auc}.

\section{Results and Discussions}
\label{sec:res}
\subsection{Line shape of $D^0D^0\pi^+$ mass spectrum}

Proceeding to the numerical analysis, we list the parameters to be fitted by the line shape data: the physical cutoff $\Lambda$, the counter terms $C_{P/V}$, and the trivial overall factor $\mathcal{N}$. It is obvious that $\Lambda$ and $C_{P/V}$ are correlated but we find that they can be strictly constrained by fitting the line shape of the $D^0D^0\pi^+$ mass spectrum. Moreover, $C_V$ has a more sensitive impact on the pole position and line shape while the dependence of $C_P$ is relatively moderate due to their coupling difference and the energy scale difference between the pseudoscalar and vector mesons in such systems. 

In order to investigate the role played by the OPE mechanism and the width effects of $D^*$, we consider three fitting schemes to fit the $D^0D^0\pi^+$ mass spectrum with Eq.~(\ref{eq:yields}), i.e.
\begin{enumerate}
    \item Scheme \rom{1}: OBE potentials excluding OPE, with a constant $D^*$ width, i.e., $\Gamma_{D^{*0}}=53.7$ keV and $\Gamma_{D^{*+}}=82.5$ keV~\cite{Albaladejo:2021vln}.
    \item Scheme \rom{2}: OBE potentials excluding OPE, with an energy-dependent $D^*$ width as formulated by Eq.~(\ref{eq:dswidth}).
    \item Scheme \rom{3}: OBE potentials with an energy-dependent $D^*$ width which incorporates with the three-body unitarity (i.e. the OPE is also properly considered).
\end{enumerate}
In the above three schemes we consider a baseline fit by fixing $C_{P/V}=0$ to compare with the fitting results with $C_V$ fixed with well-chosen values from $0.0$ to $1.0$. 

\begin{figure}[H]
    \centering
    \includegraphics[width=0.45\textwidth]{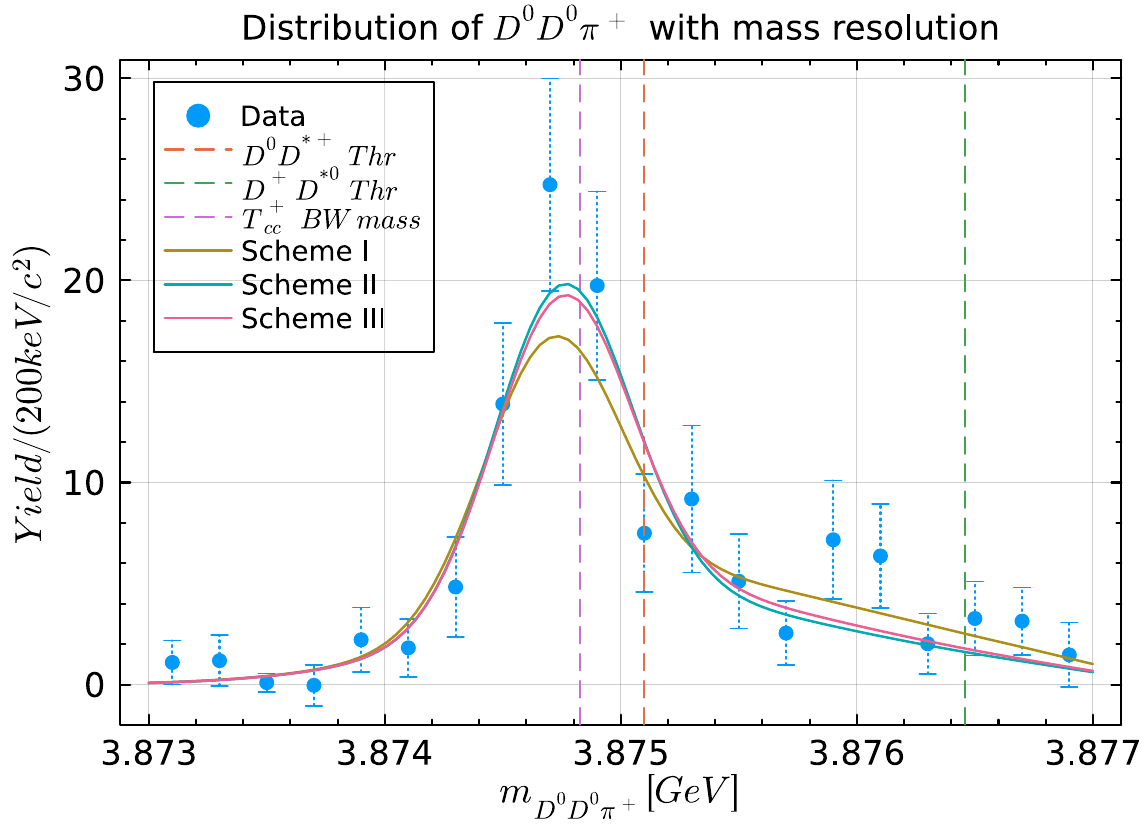}
    \caption{Baseline fitting results of line shape of $D^0D^0\pi^+$, convoluted with mass resolution function.}
    \label{fig:baselinefitting}
\end{figure}

In Table.~\ref{tab:baselinefitting} the fitted parameters (when $C_V=0.0$) and the extracted pole positions are presented. The fitted line shapes in the corresponding schemes are presented in Fig.~\ref{fig:baselinefitting}. Roughly speaking, these fitting results are comparable and reasonably good. But by detailed comparisons we can still learn some crucial information concerning the underlying dynamics:

\begin{itemize}
    \item Since the contact-range term of the OPE has been parametrized out by parameter $C_P$ which is then set as $C_P=0$, it means that the OPE will only account for the long-distance behavior of the wavefunction, while the short-distance behavior will be parametrized by $C_V$. As the consequence, the cutoff $\Lambda$ in the three schemes is consistently in the same range. We mention that if $C_P\neq 0$ the fitted values for $\Lambda$ can be very different for these three schemes. This can be compared with similar findings in  Ref.~\cite{PhysRevD.105.014024}. 
    
    \item Although the OPE contribution is subleading, it still plays an important role in fitting the line shape and extracting the pole position and width, which can be seen by comparing between Scheme-II and Scheme-III. Moreover, the width at the bound state pole including three-body unitarity (Scheme-III) is consistent with the unitarized analysis by LHCb~\cite{LHCb:2021auc}. 
    
    \item Also, it shows that the implementation of the energy-dependent $D^*$ width will reduce the pole width by a half. This indicates the importance of the three-body unitarity to some extent. 
    
\end{itemize}

\begin{figure}[H]
    \centering
    \includegraphics[width=0.45\textwidth]{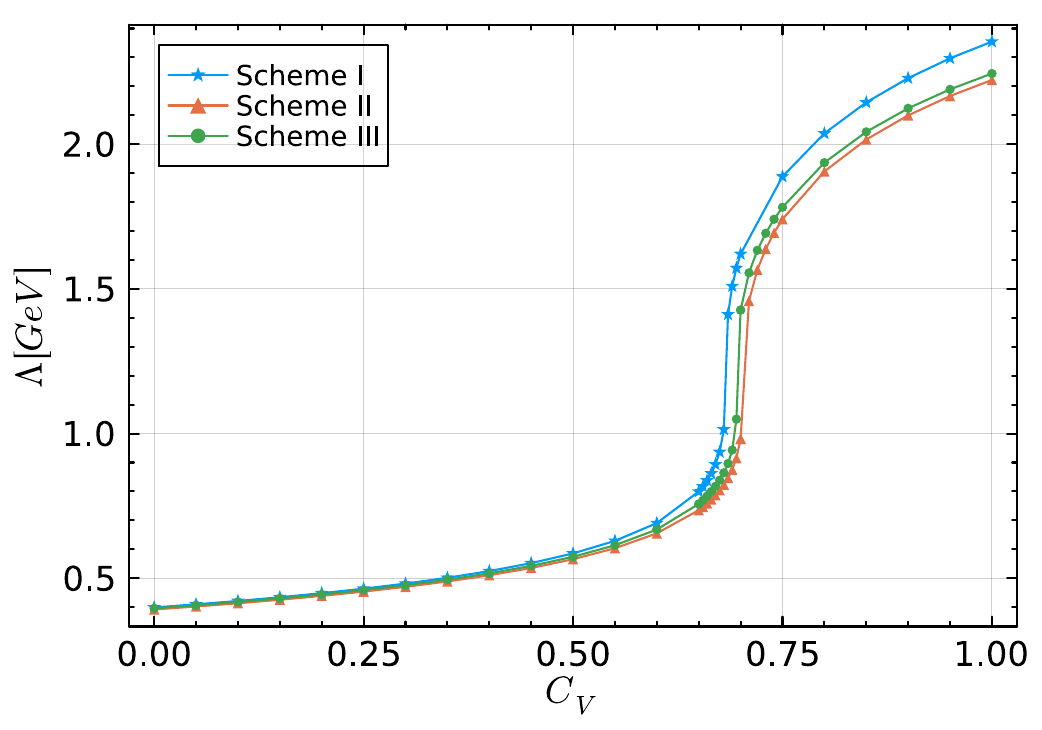}
    \includegraphics[width=0.45\textwidth]{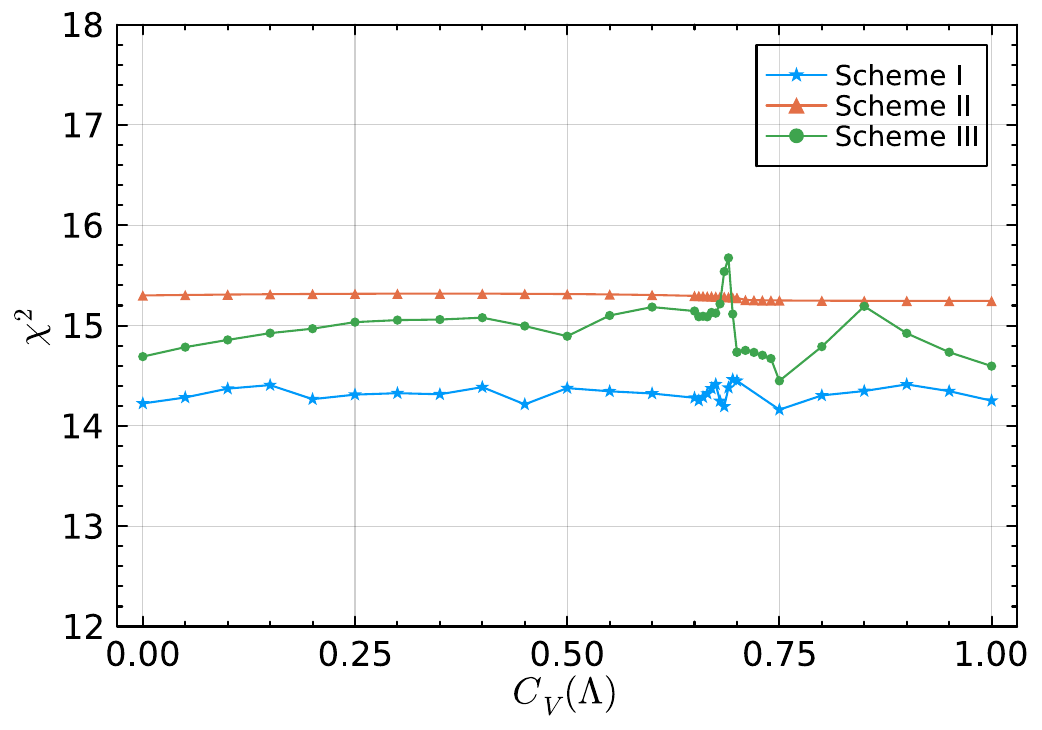}
    \caption{(Left) Dependence of $\Lambda$ on $C_V$ in these three schemes. (Right) The fluctuations of $\chi^2$ fitting on $C_V$ in these three schemes. The slight instability of $\chi^2$ in Scheme-III is caused by the mild dependence of the fitting scheme on parameter $\omega$ used in the contour deformation. But one can read from the left panel that such a fluctuation does not cause significant deviations of $\Lambda$ from its smooth correlation with $C_V$.}
    \label{fig:depcv}
\end{figure}

As mentioned earlier, there exist strong correlations between the counter terms $C_{P/V}$ and $\Lambda$. Thus, we make a survey of their correlation by looking at the variations of $\Lambda$ and $\chi^2$ with $C_P$ being fixed to zero and $C_V$ taking values within $[0,1]$. The correlation is demonstrated in Fig.~\ref{fig:depcv}. It shows that, as $C_V$ increases, the $\Lambda$ increases with different increasing rates in the small (physical) and large (unphysical) momentum regions (see Appendix.~\ref{appendix.c} for a detailed explanation). The cutoffs in the three schemes, especially in the physical region ($\leq 0.5$ GeV), are consistent with each other due to the NLO roles played by the three-body cuts.

\begin{table}[H]
\centering
  \begin{tabular}{ cccc } 
 \hline
 Schemes & $\chi^2/d.o.f$ & $\Lambda$[GeV] &$\sqrt{s}_{\text{pole}}$[keV] \\ 
 \hline
 Scheme I & $14.22/(20-1)=0.740$ & $0.399\pm 0.0008$& $-379.9^{+15.7}_{-15.5}-i\cdot 37.0^{+0.0}_{-0.0}$(RS-I) \\
 Scheme II & $15.3/(20-1)=0.805$ & $0.390\pm 0.0006$& $-347.4^{+11.5}_{-11.3}-i\cdot 18.7^{+0.2}_{-0.2}$(RS-II) \\ 
 Scheme III & $14.69/(20-1)=0.773$ & $0.396\pm 0.0010$& $-350.4^{+18.3}_{-18.2}-i\cdot 24.6^{+0.3}_{-0.2}$(RS-II)\\
\hline
\end{tabular}
\caption{Parameters and pole positions of the $T_{cc}^+$ relative to $D^0D^{*+}$ threshold. The first Riemann sheet (RS-I) in Scheme-I is defined by the two-body branch point given by Eq.~(\ref{eq:twors}) and the second Riemann sheet (RS-II) in Scheme II-III is the most important unphysical Riemann sheet accessed by Eq.~(\ref{eq:self-energy-ac}). The uncertainty of $\Lambda$ is evaluated by $\chi^2$ fitting and propagates to pole positions.}
  \label{tab:baselinefitting}
\end{table}

For isoscalar systems excluding  $1^{+-}$ $D\bar{D}^*$ and $2^{++}$ $D^*\bar{D}^*$ (see Appendix~\ref{appendix.a}), the larger $C_V$ is, the larger repulsive contact potential will be introduced and thus the larger $\Lambda$ is required to retain more short-distance potentials. Meanwhile, for a relatively large $\Lambda$, more contributions from higher-order terms, such as $\mathcal{O}(\boldsymbol{q}^2/(\boldsymbol{q}^2+\mu_{\pi}^2))$ of OPE, may manifest themselves as shown from the small but sizable deviation (due to the weak scaling of pion-exchange) between Scheme-II and Scheme-III.
Nevertheless, the reasonable range of the cutoff parameter $\Lambda$ is set empirically below $1.2$ GeV. 

Actually, the stability of the $\chi^2$ for the whole range of $C_V$ with the corresponding range of $\Lambda$ in Fig.~\ref{fig:depcv} indicates the model-independent feature in this analysis after the proper treatment of the short and long-distance dynamics and the three-body unitarity~\footnote{We mark here that the instability of $\chi^2$ in Scheme-III results from the mild dependence of the width of the pole on the parameter $\omega$, which shall be determined model-independently through the cutting diagrams by the Cutkosky rule, and the stability is anticipated if the numerical process is treated perfectly}. This also reflects the matching between the OBE and the EFT approach (see e.g. Ref.~\cite{PhysRevD.105.014024}). Such a feature can be further investigated in the study of the  $D^*D^* \ (I=0)$ system as the HQSS partners of $T_{cc}^+$. On the one hand, we expect that the HQSS relation should provide some constraints on the parameters fixed by the $T_{cc}^+$ line shape. On the other hand, we also anticipate deviations which are the manifestations of the HQSS breaking effects.

\subsection{Possible hadronic molecules in the nonstrange charmed sector}
\begin{figure}[H]
    \centering
    \includegraphics[width=0.5\textwidth]{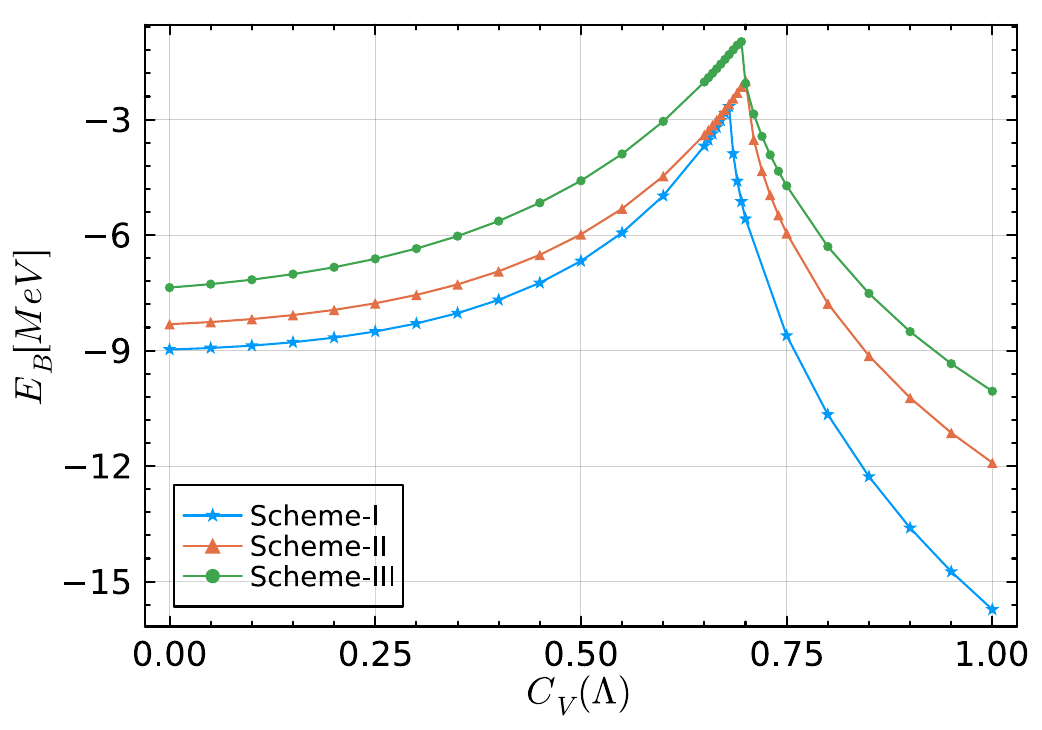}
    \caption{Dependence of the binding energy of $D^*D^*(I=0)$ on $C_V(\Lambda)$ in the three schemes. The long-distance OPE is removed in the Scheme-I and Scheme-II while retained in the Scheme-III correspondingly.}
    \label{fig:dsds}
\end{figure}

For the double-charm system, the only possible hadronic molecule except $T_{cc}^+$ is the isoscalar $1^+$ $D^*D^*$ and the contact potentials between them are strictly correlated by the heavy-quark spin symmetry. Namely, in the heavy quark limit, we have
\begin{align}
    V_{\text{CT}}(DD^*,I=0)=V_{\text{CT}}(D^*D^*,I=0) \ .
\end{align}
Thus, the parameters learned from $T_{cc}^+$ can make a strong constraint on the binding energy $E_B$ of the isoscalar $D^*D^*$. We plot the dependence of $E_B$ on the aforementioned $C_V(\Lambda)$ in the three schemes as in Fig.~\ref{fig:dsds} and the $E_B$ is found to be about $2\sim 9$ MeV for reasonable parameters ($C_V<0.7, \Lambda<1.0$ GeV), which is consistent with the result in Ref.~\cite{Dong:2021bvy} (but the isoscalar $D^*D^*$ is only found to be a bound state here).

As to the charmed-anticharmed system, the nonperturbative short-distance interactions can be very different from the double-charm ones, such as the difference between the quark annihilation of $c\bar{c}$ and quark rearrangement of $cc$. Such dynamical differences cannot be fully compensated by the meson-exchange model, and neither by an EFT approach given the HQSS breaking is unavoidable. To be consistent with the experimental results, it is found that $C_V$ tends to be much larger under the same $\Lambda$ when $C_P$ is still fixed as zero. For example, the binding energy of $D^*\bar{D}^*(I=0)$ would be found around $100$ MeV if the parameters from $T_{cc}^+$ adopted. Such a value has obviously overestimated the binding since the binding momentum $|\boldsymbol{p}|$ would amount to hundreds of MeV when its binding energy is just $10$ MeV. Meanwhile, it will be questionable for the Born approximation since it would not hold for such deeply bound states. On the other hand, taking into account that the isoscalar $D\bar{D}^*$ with $C=+$ is found to form a bound state for a large $C_V$, but only a virtual state for a small $C_V$, the existence of $X(3872)$ allows us to limit our discussions within $0.35< \Lambda< 0.65$ GeV and a relatively large $0.7< C_V< 1.0$ below though we present the results within a wide range of $C_V$ in Fig.~\ref{fig:dsds} to show the $\Lambda$ dependence~\footnote{As shown later, the results calculated with $\Lambda\sim 0.4$ GeV are more consistent with the observable. Otherwise, the value of $C_V$ is required to be larger than unit which corresponds to larger $\Lambda$ and will become unphysical. }. The LSE solutions for the isoscalar $D^{(*)}\bar{D}^{(*)}$ are listed in Tab.~\ref{tab:twoccbar}.

\begin{table}[H]
\centering
  \begin{tabular}{ cccc } 
 \hline
 System & $J^{PC}$& Pole($\Lambda=0.35$GeV)&Pole($\Lambda=0.65$GeV)\\ 
 \hline
 $D\bar{D}$& $0^{++}$&$(1,0.3)$& $(1,33.6)$ \\ 
 \hline
 \multirow{3}{*}{$D^*\bar{D}^*$}&$0^{++}$&$(1,0.006)-(1,12.4)$&$(1,0.5)-(1,39.3)$\\
  &$1^{+-}$ &$(1,0.4)-(1,6.4)$ &$(1,4.1)-(1,37.7)$\\
  &$2^{++}$ &$(2,3.3)-(1,3.2)$ &$(1,46.6)-(1,117.7)$\\
 \hline
\end{tabular}
\caption{Pole positions (in MeV) of the isoscalar $D^{(*)}\bar{D}^{(*)}$. The notation $(rs,E_B)$ represents the binding energy $E_B$ on the first/second Riemann sheet. The $E_B$ boundary for each $\Lambda$ is evaluated with $C_V=0.7, \ 1.0$, respectively. See the context for further explanations.}
  \label{tab:twoccbar}
\end{table}

Since there is no spin-dependent terms for $D\bar{D}$, the above $\Lambda$ boundary is chosen partly by comparison with another calculation by dimensional regularization (DR) in Ref.~\cite{Dong:2021bvy}. While for $D^*\bar{D}^*$, there is a mass splitting for different $J$ due to the inclusion of spin-dependent terms and we note that the unacceptably large numerical binding energy $117.7$ MeV for the $2^{++}$ $D^*\bar{D}^*$ system is calculated with the unphysical value of $C_V=1.0$. In such a case the full reservation of the Dirac Delta term will introduce an extremely large attractive potential as mentioned in the last subsection. Though more profound considerations, such as higher partial waves, coupled-channel effect, and four-body unitarity, etc, are needed, the above result shows the signal for the existence of the isoscalar hadronic molecules composed of $D^*\bar{D}^*$, especially the $2^{++}$ tensor state. Despite the fact that there is no lower channels than the $D\bar{D}$ threshold in strong decays, the existence of an isoscalar $D\bar{D}$ has been widely predicted by phenomenological studies~\cite{Gamermann:2006nm,Nieves:2012tt} and by LQCD calculation~\cite{Prelovsek:2020eiw}. Besides, the role of the isoscalar $D\bar{D}$ state can be significant in the study of its HQSS partners like $X(3872)$ by the loop enhancement if there exists a near-threshold bound/virtual state~\cite{Guo:2014hqa,Dai:2019hrf}. As for the isovector $D^{(*)}\bar{D}^{(*)}$, we found no bound/virtual states here but the observed hadronic molecule candidate $Z_c(4020)^{\pm}$~\cite{BESIII:2013mhi,BESIII:2013ouc} is actually generated by more complicated mechanisms, similar to $Z_c(3900)$ to be discussed later.

\begin{figure}[H]
    \centering
    \includegraphics[width=0.8\textwidth]{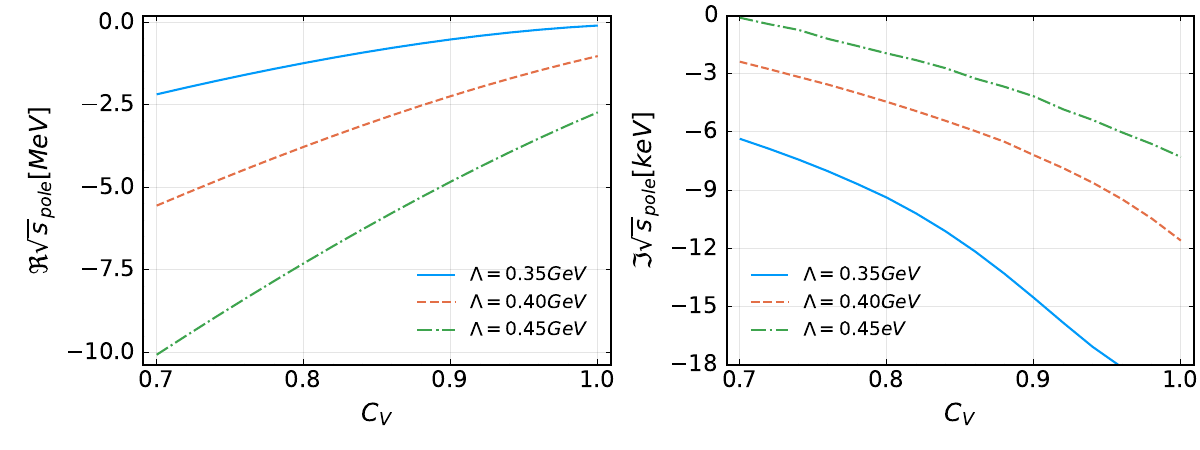}
    \includegraphics[width=0.8\textwidth]{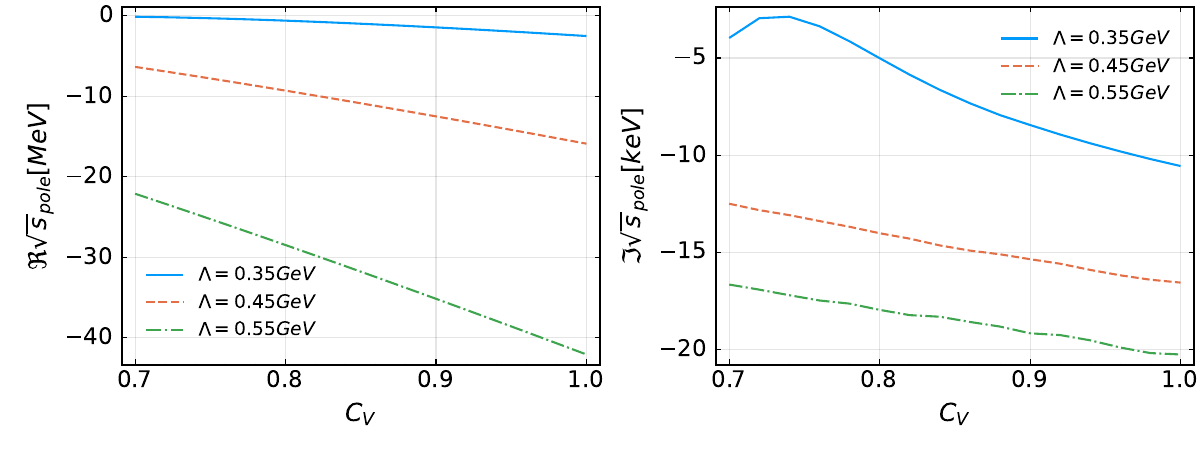}
    \caption{Real and imaginary parts of pole positions relative to the $D^0\bar{D}^{*0}+c.c$ threshold of isoscalar $D\bar{D}^*$ with the positive (upper panels) and negative (lower panels) $C$-parity.
    } 
    \label{fig:eb_ccbar}
\end{figure}

For the $D\bar{D}^*$ system, the three-body unitarity has to be taken into account. Interestingly, it seems that there should exist a counterpart of $X(3872)$ with negative $C$-parity, denoted as $\tilde{X}(3872)$, which has been studied in the literature~\cite{COMPASS:2017wql,Chen:2010ze,Wang:2020dgr,Ortega:2021yis}.  In Fig.~\ref{fig:eb_ccbar}, we plot the pole positions of these two states with respect to $\Lambda$ and $C_V$. Similarly, the binding energy of $X(3872)$ is estimated to be from hundreds of keV to a few MeV along with an imaginary part comparable with that of $T_{cc}^+$. One notices the different dependence behaviors of the width and binding energy $E_B$ on the short-distance parameter $C_V$ for $X(3872)$. Namely, with the increase of $C_V$ the value of $|E_B|$ drops and the width (defined as twice of the absolute value of the imaginary part of the pole position) increases. In particular, with $C_V\to 1$, the width increases up to $40\sim 60$ keV where $E_B$ is only hundreds of keV. This behavior is consistent with the results of Refs.~\cite{Fleming:2007rp,Baru:2011rs}. In the upper panels of Fig.~\ref{fig:eb_ccbar} one also sees the correlation of the width and binding energy with the cutoff parameter $\Lambda$ with the fixed $C_V$. With the smaller value for $\Lambda$, one obtains a relatively shallower binding and relatively larger width.  Note that with the fixed $C_V$ the smaller value of $\Lambda$ corresponds to the smaller contributions from the short-distance dynamics. The driving mechanism for the binding of $X(3872)$ due to the short-distance dynamics is actually highlighted.

In contrast with $X(3872)$, the $C=-1$ isoscalar $\tilde{X}(3872)$ is found to be more bound with a slightly smaller imaginary part. Actually, if one only looks at the solid lines for both $X(3872)$ and $\tilde{X}(3872)$ in Fig.~\ref{fig:eb_ccbar}, their values are quite close to each other. However, with the increase of the $\Lambda$ value the binding energy and total width of $\tilde{X}(3872)$ both increase. As mentioned earlier that the larger value of $\Lambda$ corresponds to larger contributions from the short-distance dynamics, it suggests that the coupling strength for $\tilde{X}(3872)$ to $D\bar{D}^*+c.c.$ increases. Such a change, on the one hand, leads to a larger binding energy, and on the other hand, overtakes the decrease of the phase space to produce a larger decay width. With  $\Lambda\sim 0.4$ GeV favoured by $X(3872)$ (by comparing the extracted pole mass with the value from the Particle Data Group (PDG)~\cite{Workman:2022ynf}), the binding energy of $\tilde{X}(3872)$ is about 10 MeV, which is very close to the result $M_{\tilde{X}(3872)}=3860.4\pm 10.0 \ \text{MeV}/c^2$ reported by the COMPASS Collaboration~\cite{COMPASS:2017wql}. However, it should be cautioned that for such a deeply bound state, the formula of the $D^*$ width must be revised regarding the small mass difference between $D^{*}$ and $D\pi$, and the large binding energy up to 10 MeV. To be brief, comparing the behaviors of $X(3872)$ and $\tilde{X}(3872)$ with each other one sees the crucial role played by the short-distance dynamics. Meanwhile, one sees the apparent deviations from the HQSS.

For the isovector $D\bar{D}^*$ system, the interactions from the $\rho$ and $\omega$ meson exchanges actually cancel with each other. Without additional mechanisms this system cannot form bound state through the residual meson-exchange potentials (i.e. $\sigma$, $\pi$, $\eta$, $J/\psi$, etc), and it turns out more likely to be a virtual state or resonance (see e.g. Refs.~\cite{Aceti:2014uea,Prelovsek:2013xba,HALQCD:2016ofq,Chen:2014afa,Liu:2019gmh} and the review of Ref.~\cite{Guo:2017jvc}).  This can be regarded as being consistent with the experimental observations of $Z_c(3900)$~\cite{BESIII:2013ris,Belle:2013yex,BESIII:2013qmu}. In particular, the observation of $Z_c(3900)$ seems to be strongly correlated with the production of $Y(4260)$ for which the presence of the so-called ``triangle singularity" can provide a natural explanation for its enhanced production rate~\cite{Wang:2013cya,Wang:2013hga}.

\section{Summary}
\label{sec:con}
Motivated by the observation of the double-charm tetraquark molecule candidate $T_{cc}^+$ by the LHCb Collaboration, we present a coupled-channel analysis of the $D^{0}D^{0}\pi^+$ mass spectrum within the meson-exchange model including isospin breaking effect. It is found that the dynamical details, and the two-body or three-body unitarity, have indeed played an important role in the description of the underlying dynamics for such composite states with unstable constituents. The well-described line shape allows us to extract the pole information about $T_{cc}^+$ at a rather high precision and it confirms the nature of $T_{cc}^+$ as an isoscalar $DD^*$ hadronic molecule which can be regarded as a milestone in the study of hadronic molecules and hadron spectroscopy. Moreover, by implementing the HQSS it provides a stringent constraint on some of those crucial dynamical aspects for the $D^{(*)}\bar{D}^{(*)}$ systems which allows us to compare with the EFT results and examine the HQSS breaking effects through the combined analysis. In particular, the existence of another double-charm hadronic molecule candidate $D^*D^*(I=0)$ is predicted in the same framework. 

In the combined analysis the interactions between the charmed mesons are hypothesized to be saturated by the meson-exchange potentials. This allows us to relate the OBE potentials for the double-charm system with the charmed-anticharmed system. Although it should be recognized that the short-distance dynamics between the charm quark rearrangement and charm-anticharm annihilations are actually different, we expect that these two systems will still share some crucial aspects of the dynamics via the OBE potentials. It allows us to establish the relations between the double-charm and charmed-anticharmed systems, and extract the binding conditions for these systems, especially for the isoscalar ones. For isovector states such as $Z_c(3900)$ and $Z_c(4020)$, we find that the light vector mesons potentials via the exchange of $\rho^0$ and $\omega$ actually cancel each other. The residual potentials turn out to be insufficient for binding. It suggests that $Z_c(3900)$ and $Z_c(4020)$ should behave more likely to be threshold-enhanced virtual states, resonances, or even quasi-bound states depending on additional mechanisms introduced. Besides, the correlation of the $Z_c(3900)$ and $Z_c(4020)$ productions with the triangle singularity mechanism make it necessary to include the production mechanisms in the detailed analysis. While these are still non-trivial issues to be addressed, we leave them to be studied in future works.

\section*{Acknowledgement}
The authors thank Meng-Chuan Du and Yin Cheng for their early help on this work. Also the authors thank Ying Chen, Meng-Lin Du, Feng-Kun Guo, Xiao-Hai Liu, Peng-Yu Niu, Qian Wang, and Jia-Jun Wu for their useful discussions. This work is supported, in part, by the National Key Basic Research Program of China under Contract No. 2020YFA0406300, National Natural Science Foundation of China (Grant No. 12235018),  DFG and NSFC funds to the Sino-German CRC 110 ``Symmetries and the Emergence of Structure in QCD'' (NSFC Grant No. 12070131001, DFG Project-ID 196253076), and the Strategic Priority Research Program of Chinese Academy of Sciences (Grant No. XDB34030302).

\appendix 
\section{Wave functions and meson-exchange potentials}
\label{appendix.a}

For the $S$-wave $DD$ ($D^*D^*$) systems, the isoscalar and isovector wave functions are $[D^{(*)}D^{(*)}]_{\mp}=-\frac{1}{\sqrt{2}}(D^{(*)0}D^{(*)+}\mp D^{(*)+}D^{(*)0})$. However, some of those are forbidden due to the Bose-Einstein statistic: $L+S+I+2i=\text{even number}$, with $L$ and $S$ the total orbital angular momentum and spin of the two mesons; $I$ and $i$ the total and individual isospin of the mesons (see Eq.~(15) in Ref.~\cite{Abreu:2022sra}). For instance, for the $DD$ system, the $I=0$ state is forbidden. The meson-exchange potentials are,
\begin{align}
    V^{[DD]_{+}}&=-g_S^2\tilde{\mathcal{X}}'_{\sigma}-\frac{g_V^2\beta^2}{4}\mathcal{X}'_{\rho^0}+\frac{g_V^2\beta^2}{4}\mathcal{X}'_{\omega}+\frac{g_V^2\beta^2}{2}\mathcal{X}'_{\rho^-} \ ,\\
    V^{[D^*D^*]_{\mp}}&=-g_S^2\tilde{\mathcal{X}}''_{\sigma}-\frac{1}{2}(\frac{g}{f_{\pi}})^2\mathcal{Y}''_{\pi^0}+\frac{1}{6}(\frac{g}{f_{\pi}})^2\mathcal{Y}''_{\eta}-\frac{g_V^2\beta^2}{4}\mathcal{X}''_{\rho^0}+2\lambda^2g_V^2\mathcal{Z}''_{\rho^0}+\frac{g_V^2\beta^2}{4}\mathcal{X}''_{\omega}-2\lambda^2g_V^2\mathcal{Z}''_{\omega} \ ,\\
    &+\frac{g_V^2\beta^2}{2}\mathcal{X}''_{J/\psi}-4\lambda^2g_V^2\mathcal{Z}''_{J/\psi}+(\frac{g}{f_{\pi}})^2\mathcal{Y}''_{\eta_c}\mp [\frac{g_V^2\beta^2}{2}\mathcal{X}''_{\rho^+}-4\lambda^2g_V^2\mathcal{Z}''_{\rho^+}+(\frac{g}{f_{\pi}})^2\mathcal{Y}''_{\pi^+}] \ ,
\end{align}
where we have defined some new functions,
\begin{align}
        \tilde{\mathcal{X}}'_{\text{ex}}&=\frac{1}{|\boldsymbol{q}|^2+m_{\text{ex}}^2-{q^{0}}^2}, &
        \mathcal{X}'_{\text{ex}}&=\frac{(1-\frac{{q^0}^2}{m_{\text{ex}}^2})}{|\boldsymbol{q}|^2+m_{\text{ex}}^2-{q^{0}}^2} \ ,\\
        \tilde{\mathcal{X}}''_{\text{ex}}&=\frac{\boldsymbol{\epsilon}_1\cdot\boldsymbol{\epsilon}_3^*\boldsymbol{\epsilon}_2\cdot\boldsymbol{\epsilon}_4^*}{|\boldsymbol{q}|^2+m_{\text{ex}}^2-{q^{0}}^2}, &
        \mathcal{X}''_{\text{ex}}&=\frac{(1-\frac{{q^0}^2}{m_{\text{ex}}^2})\boldsymbol{\epsilon}_1\cdot\boldsymbol{\epsilon}_3^*\boldsymbol{\epsilon}_2\cdot\boldsymbol{\epsilon}_4^*}{|\boldsymbol{q}|^2+m_{\text{ex}}^2-{q^{0}}^2} \ ,\\
        \mathcal{Y}''_{\text{ex}}&=\frac{(\boldsymbol{\epsilon}_1\times\boldsymbol{\epsilon}_3^*)\cdot\boldsymbol{q}(\boldsymbol{\epsilon}_2\times\boldsymbol{\epsilon}_4^*)\cdot\boldsymbol{q}}{|\boldsymbol{q}|^2+m_{\text{ex}}^2-{q^{0}}^2},&
        \mathcal{Z}''_{\text{ex}}&=\frac{(\boldsymbol{\epsilon}_1\times\boldsymbol{\epsilon}_3^*\times\boldsymbol{q})\cdot(\boldsymbol{\epsilon}_2\times\boldsymbol{\epsilon}_4^*\times\boldsymbol{q})}{|\boldsymbol{q}|^2+m_{\text{ex}}^2-{q^{0}}^2} \ ,
\end{align}
with $\boldsymbol{\epsilon}_i$ the polarization vector of initial or final $D^*$ mesons. And similarly, we make the following substitutions for the spin-dependent terms,
\begin{align}
    \mathcal{Y}''_{P/V}&\sim \frac{1}{3}(-\frac{\mu_{P/V}^2}{|\boldsymbol{q}|^2+\mu_{P/V}^2}+C_{P/V})\hat{S}(\boldsymbol{\epsilon}_1\times\boldsymbol{\epsilon}_3^*,\boldsymbol{\epsilon}_2\times\boldsymbol{\epsilon}_4^*) \ ,\\
    \mathcal{Z}''_{V}&\sim \frac{2}{3}(-\frac{\mu_V^2}{|\boldsymbol{q}|^2+\mu_V^2}+C_V)\hat{S}(\boldsymbol{\epsilon}_1\times\boldsymbol{\epsilon}_3^*,\boldsymbol{\epsilon}_2\times\boldsymbol{\epsilon}_4^*) \ .
\end{align}
where the matrix elements of operator $\hat{S}(\boldsymbol{\epsilon}_1\times\boldsymbol{\epsilon}_3^*,\boldsymbol{\epsilon}_2\times\boldsymbol{\epsilon}_4^*)$ are $2,1,-1$ for different total spin $J=0,1,2$, respectively, and the tensor part $\hat{T}(\boldsymbol{\epsilon}_1\times\boldsymbol{\epsilon}_3^*,\boldsymbol{\epsilon}_2\times\boldsymbol{\epsilon}_4^*)$ vanishes (see Appendix A in Ref.~\cite{Wang:2021aql}).

For $D\bar{D}$ ($D^*\bar{D}^*$), the wave functions are correlated with the charge conjugation parity $C=(-1)^{L+S}$  for the charge-neutral systems. Thus, the allowed quantum numbers for $D\bar{D}$ are only those with $C=+$, i.e.,  $[D\bar{D}]_{\pm}=\frac{1}{\sqrt{2}}(D^{0}\bar{D}^{0}\pm D^{+}D^{-})$ where ``$+$" corresponds to the isoscalar and ``$-$" to the isovector:
\begin{enumerate}
    \item $D^0\bar{D}^0\rightarrow D^0\bar{D}^0$/$D^+D^-\rightarrow D^+D^-$
    \begin{align}
        \begin{split}
            V^{[D\bar{D}]}&=-g_S^2\tilde{\mathcal{X}}'_{\sigma}-\frac{g_V^2\beta^2}{4}\mathcal{X}'_{\rho^0}-\frac{g_V^2\beta^2}{4}\mathcal{X}'_{\omega}-\frac{g_V^2\beta^2}{2}\mathcal{X}'_{J/\psi} \ .
        \end{split}
    \end{align}
    \item $D^0\bar{D}^0\rightarrow D^+D^-$
    \begin{align}
        V^{[D\bar{D}]}=-\frac{g_V^2\beta^2}{2}\mathcal{X}'_{\rho^-} \ .
    \end{align}
\end{enumerate}
The allowed quantum numbers for the $D^*\bar{D}^*$ system are those with $C=+,J=0,2$ or $C=-,J=1$ for wave functions $[D^{*}\bar{D}^{*}]_{\pm}=\frac{1}{\sqrt{2}}(D^{*0}\bar{D}^{*0}\pm D^{*+}D^{*-})$:
\begin{enumerate}
    \item $D^{*0}\bar{D}^{*0}\rightarrow D^{*0}\bar{D}^{*0}$/$D^{*+}D^{*-}\rightarrow D^{*+}D^{*-}$
    \begin{align}
        V^{[D^*\bar{D}^*]}&=-g_S^2\tilde{\mathcal{X}}''_{\sigma}-\frac{g_V^2\beta^2}{4}\mathcal{X}''_{\rho^0}+2\lambda^2g_V^2\mathcal{Z}''_{\rho^0}-\frac{g_V^2\beta^2}{4}\mathcal{X}''_{\omega}+2\lambda^2g_V^2\mathcal{Z}''_{\omega}-\frac{g_V^2\beta^2}{2}\mathcal{X}''_{J/\psi}+4\lambda^2g_V^2\mathcal{Z}''_{J/\psi}\\
        &+\frac{1}{2}(\frac{g}{f_{\pi}})^2\mathcal{Y}''_{\pi^0}+\frac{1}{6}(\frac{g}{f_{\pi}})^2\mathcal{Y}''_{\eta}+(\frac{g}{f_{\pi}})^2\mathcal{Y}''_{\eta_c} \ .
    \end{align}
    \item $D^{*0}\bar{D}^{*0}\rightarrow D^{*+}D^{*-}$
    \begin{align}
        V^{[D^*\bar{D}^*]}&=-\frac{g_V^2\beta^2}{2}\mathcal{X}''_{\rho^-}+4\lambda^2g_V^2\mathcal{Z}''_{\rho^-}+(\frac{g}{f_{\pi}})^2\mathcal{Y}''_{\pi^-} \ .
    \end{align}
\end{enumerate}

\section{Analytic structures}
\label{appendix.b}
We present a detailed treatment for the branch points/cuts and analytic continuation encountered when solving the coupled-channel LSE. For the $^3S_1$ system, Eq.~(\ref{eq:lse}) can be reduced to,
\begin{align}
    \begin{split}
         T_{\alpha\beta}(p,k;E)&=V_{\alpha\beta}(p,k;E)+\sum_{\delta}\int_{0}^{\Lambda}\frac{dqq^2}{2\pi^2}V_{\alpha\delta}(p,q;E)G_{\delta\delta}(q;E)T_{\delta\beta}(q,k;E) \ .
    \end{split}
\end{align}
Note that we have used the following relations for the projected partial-wave potentials $V_{ij}$ in the above equation~\cite{Oller:2018zts} ,
\begin{align}
    \begin{split}
        V_{ij}(-p_1,p_2)&=(-1)^{l_i}V_{ij}(p_1,p_2) \ ,\\
        V_{ij}(p_1,-p_2)&=(-1)^{l_j}V_{ij}(p_1,p_2) \ .
    \end{split}
\end{align}
with $l_i$ the angular momentum and $p_{1,2}$ the initial/final magnitude of the three-vector momentum.

As discussed in Refs.~\cite{Oller:2018zts,PhysRevC.84.015205,Ji:2022blw}, the most important branch points/cuts come from two kinds: the RHCs induced by the thresholds (open channels) and the left-hand cuts (LHCs) arising from the particle-exchange potentials.

\subsection{The participation of new channel}
For each channel, there will be a new branch point located at the threshold which leads to a new branch cut (usually starting from the threshold to the positive infinity). For a channel with stable particles, e.g. $\pi\pi$, the branch point is located at $2m_{\pi}$ on the physical axis and thus they usually have a significant influence on the observable. For a channel with an unstable particle, e.g. $\sigma N$,  the branch points are located at $m_N+m_{\sigma}\pm i\Gamma_{\sigma}/2$ where $\Gamma_{\sigma}$ is the width of  $\sigma$. Since they are always on the complex plane, the poles on these Riemann Sheets induced by these cuts are of no interests in most cases.

\begin{figure}[H]
    \centering
    \includegraphics[width=15cm,scale=0.5]{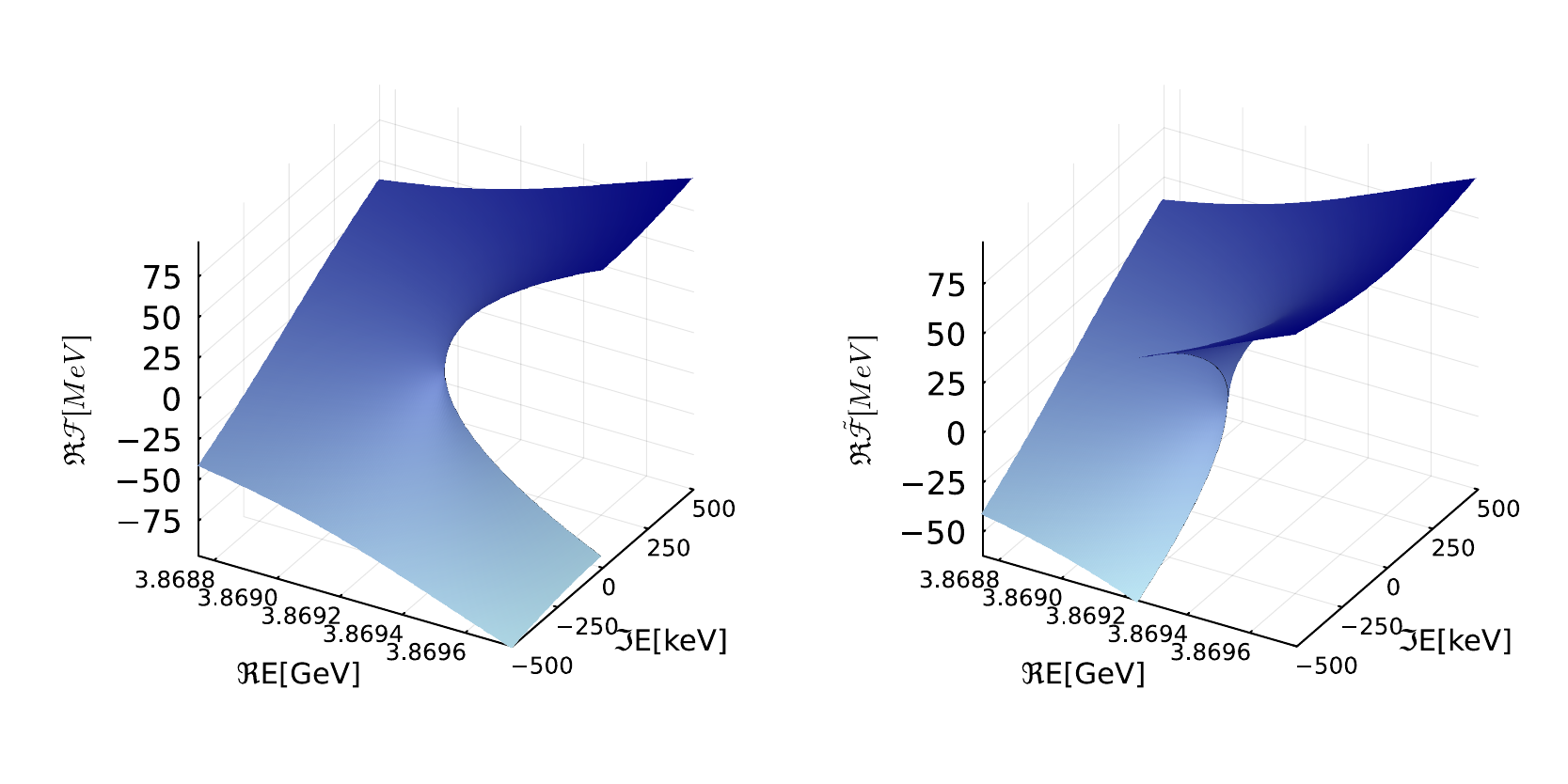}
    \caption{The real part of $p_{cm}$ of $D^{*+}[D^0\pi^+]D^0$ before(a)/after(b) the analytic continuation within the energy range $-500 \ \text{keV}\leq \Re E\leq 500 \ \text{keV}$ relative to $D^0\pi^+D^0$ threshold and width range $-500 \ \text{keV}\leq\Im E\leq 500 \ \text{keV}$ when the frame $D^{*+}[D^0\pi^+]$ is static, i.e., $\boldsymbol{q}=\boldsymbol{0}$. The three-body cut is redirected from the RHC on the real axis to the direction parallel to the negative imaginary axis.}
    \label{fig:self-energy-ac}
\end{figure}

Therefore, for $DD^*\rightarrow DD^*$ process, the most relevant branch points are those $DD\pi$ thresholds ($D^0D^0\pi^+$, $D^0D^+\pi^0$ and $D^+D^+\pi^-$)~\cite{PhysRevC.84.015205}. For the case of $\Im E>0$, the path of $E'$ (transformed by Eq.~(\ref{eq:boost})) when integrating $q$ from zero to positive infinity along the real axis (undeformed path) does not cross the cut (see Fig.~7 in Ref.~\cite{Doring:2009yv}) and hence there is no analytic problem. However, when $E$ enters into the lower half plane, i.e. $\Im E<0$ for $\Re E>m_1+m_2+m_3$, the corresponding undeformed path in $E'$ complex plane will enclose the RHC and thus leads to discontinuity. To analytically continue to the lower half plane of $E$, the branch cut of $D^*[D\pi]$ self-energy can be redirected along the negative imaginary axis which means the function $\mathcal{F}(E,\boldsymbol{q},m_1,m_2,m_3)$ is replaced by $\tilde{\mathcal{F}}(E,\boldsymbol{q},m_1,m_2,m_3)$ in Eq.~(\ref{eq:self-energy-ac}) first. We plot the $\mathcal{F}(E,\boldsymbol{0},m_{D^{0}},m_{\pi^+},m_{D^0})$ and $\tilde{\mathcal{F}}(E,\boldsymbol{0},m_{D^{0}},m_{\pi^+},m_{D^0})$ as a demonstration in Fig.~\ref{fig:self-energy-ac}.

Meanwhile, the integral path of ${q}$ shall be deformed and we adopt the parametrization used in Ref.~\cite{Sadasivan:2021emk}, namely,
\begin{equation}
    \Gamma_{\text{SMC}}=\{q|q=t+iV_0(1-e^{-t/w})(1-e^{(t-\Lambda)/w}),t\in [0,\Lambda]\}
\end{equation}

With $\omega\sim0.1$ GeV and $V_0\sim-0.1$ GeV, the ``spectator momentum contour" (SMC) can be deformed and deep into the lower half complex plane with the endpoints $q=0,\Lambda$ fixed. And then the deformed path of $E'$ can avoid crossing the branch cut parallel to the negative imaginary axis, given by $\tilde{\mathcal{F}}$ (see Fig. 7 in Ref.~\cite{Doring:2009yv}). Another advantage of such a parametrization is that the divergence of the pion-exchange potentials can also be moderated to some extent~\footnote{This is understandable since the three-body cuts enter into two aspects, i.e. the $D\pi$ self-energy and the on-shell OPE.} which we will discuss in the next subsection.

\subsection{The particle-exchange potentials}
Due to the unitarity, the $t$ and $u$-channel particle-exchange potentials can always lead to the LHCs when projecting these potentials into partial waves. For a general process $a(\boldsymbol{p}_1)b(\boldsymbol{p}_2)\rightarrow c(\boldsymbol{p}_3)d(\boldsymbol{p}_4)$, the $t$-channel $S$-wave projected potential is,
\begin{align}
\label{eq:tcuts}
    \begin{split}
        &\int_{-1}^{1}\frac{dz}{t-m_{\text{ex}}^2}\\
        &=\int_{-1}^{1}\frac{dz}{(E_3-E_1)^2-p_1^2-p_3^2+2 z p_1 p_3-m_{\text{ex}}^2}\\
        &=\frac{1}{2p_1p_3}\log(\frac{(p_3-p_1)^2-(E_3-E_1)^2+m_{\text{ex}}^2}{(p_3+p_1)^2-(E_3-E_1)^2+m_{\text{ex}}^2})\\
        &=\frac{1}{2p_1p_3}[\log((p_3-p_1)^2-(E_3-E_1)^2+m_{\text{ex}}^2)-\log((p_3+p_1)^2-(E_3-E_1)^2+m_{\text{ex}}^2)]
    \end{split}
\end{align}
with $m_{i}$/$E_i$/$p_i$ ($i=1, \ 2, \ 3, \ 4$) the mass/energy/module of the three-vector momentum of particle $a,b,c,d$ in order, and $m_{\text{ex}}$ the mass of the exchanged particle. Note that the last second step to the last one is reasonable only for $p_i$ on the physical axis or around it. Otherwise, the potential in such a form has to be analytically extrapolated properly. The branch cuts are thus determined by restricting the arguments of the two logarithms into $(-\infty,0)$, i.e.,
\begin{equation}
    \label{eq:dcs}
    p_3=\pm p_1\pm\sqrt{(E_3-E_1)^2-m_{\text{ex}}^2-x^2},\quad x\in R \ ,
\end{equation}
where the two ``$\pm$" signs are uncorrelated. Equation~(\ref{eq:dcs}) gives rise to the curved cut, called dynamical cuts (DCs), with $p_3$ as a function of $p_1$ and the exact discontinuities can be evaluated by the N/D method~\cite{Oller:2018zts}. As for $t$-channel $|E_3-E_1|\ll m_{\text{ex}}$ in our case $DD^*\rightarrow DD^*$($D\bar{D}^*\rightarrow D\bar{D}^*$), the on-shell, half-off-shell and off-shell projected partial potentials are argued as follows: 
\begin{enumerate}
    \item For the on-shell case $p_1\approx p_3$, Eq.~(\ref{eq:dcs}) with minus sign in the first $\pm$ symbol gives,
    \begin{equation}
        p_3\approx\pm\frac{1}{2}\sqrt{(E_3-E_1)^2-m_{\text{ex}}^2-x^2} \ ,
    \end{equation}
    which leads to the LHC along $p_3^2\leq-\frac{m_{\text{ex}}^2}{4}$, which means that the crossing of cuts only happens when $|\Im p_3|\geq\frac{m_{\text{ex}}}{2}$.
    \item For the half-off-shell case $p_1>0$, the imaginary part of $p_3$, dominated by $m_{\text{ex}}$, is still far from the physical region of interests (about tens of MeV mostly) since the masses of the $t$-channel exchanged mesons are usually at the amount of hundreds of MeV ($\rho$, $\omega$, $\sigma$, etc).
    \item For the off-shell case with $p_1,p_3>0$, there is no crossing of cuts as seen from the second line in Eq.~(\ref{eq:tcuts}).
\end{enumerate}

While for the $u$-channel processes (by performing $p_3\leftrightarrow p_4$), the situation turns to be subtle as $|E_4-E_1|$ becomes comparable with or even larger than $m_{\text{ex}}$ for the exchanged pion while still the same for heavier exchanged mesons. Similarly, the projected partial potentials are treated as follows:
\begin{enumerate}
    \item For the on-shell case, the poles of the integrand $(E_4-E_1)^2-p_4^2-p_1^2+2 z p_1p_4-m_{\pi}^2=0$ for $z$ within the interval $[-1,1]$ actually lead to different branch cuts for the three different processes involved: 1) a finite cut on the real axis above the $D^0D^{*+}$ threshold for process $D^{0}D^{*+}\xrightarrow{\pi^{+}}D^{*+}D^{0}$; 2) a finite cut along with a circular cut as shown in Fig.~22 in Ref.~\cite{Doring:2009yv} for process $D^{0}D^{*+}\xrightarrow{\pi^{0}}D^{*0}D^{+}$; 3) a finite cut crossing the $D^0D^{*+}$ threshold for $D^{+}D^{*0}\xrightarrow{\pi^{+}}D^{*0}D^{+}$ which can be treated analytically as shown later.
    \item For the off-shell case with $p_1,p_4>0$, the above condition turns out to be a finite cut $(0,\frac{|m_4^2-m_3^2-m_1^2+m_2^2|}{2m_{\pi}}]$ in the $\sqrt{s}$ complex plane (which fully covers $(0,m_{D^0}+m_{D^{*+}}]$ for process $D^{0}D^{*+}\xrightarrow{\pi^{+}}D^{*+}D^{0}$) and thus crossing over the cut will occur which produces discontinuity when the energy $\sqrt{s}$ moves from the upper half plane to the lower one. Thus, one needs to deform the integral path rather than $[-1,1]$ on the real axis such as a polyline $-1\rightarrow -1-ia\rightarrow 1-ia\rightarrow 1$ with sufficiently large $a$ as used in Ref.~\cite{Ji:2022blw} or analytically continue by adding the discontinuity on the real axis directly, i.e., twice of the imaginary part on the cut,
    \begin{equation}
    \label{eq:acope}
  \int_{-1}^{1}\frac{1}{z-\xi} = \left\{\right.
  \begin{array}{l}
   \log(1-\xi)-\log(-1-\xi)+2\pi i, \ \ \ |\Re\xi|<1\text{ and }\Im\xi<0 \ ,\\
   \log(1-\xi)-\log(-1-\xi), \ \ \ \text{else} \ ,
  \end{array}
    \end{equation}
    with $\xi=\frac{p_1^2+p_4^2+m_{\pi}^2-(E_4-E_1)^2}{2p_1p_4}$.
    \item The situation becomes even more complicated for the half-off-shell case. But the procedure goes the same as the previous case with the mentioned adjustment above.
\end{enumerate}

\begin{figure}[H]
    \centering
    \includegraphics[width=0.45\textwidth]{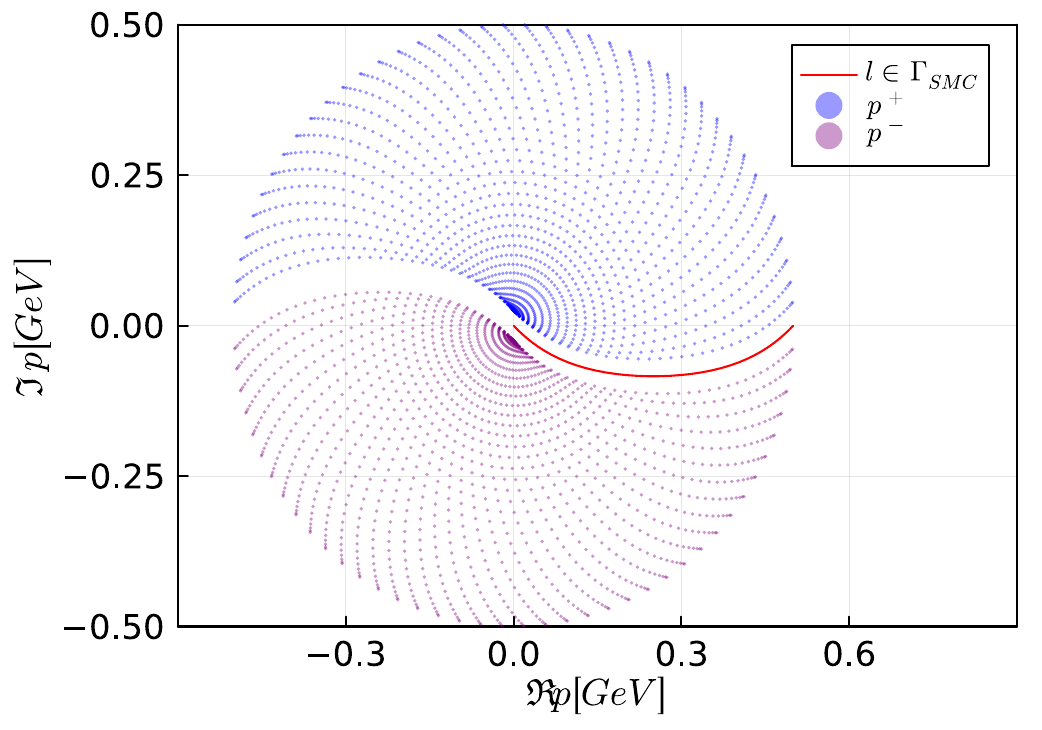}
    \includegraphics[width=0.45\textwidth]{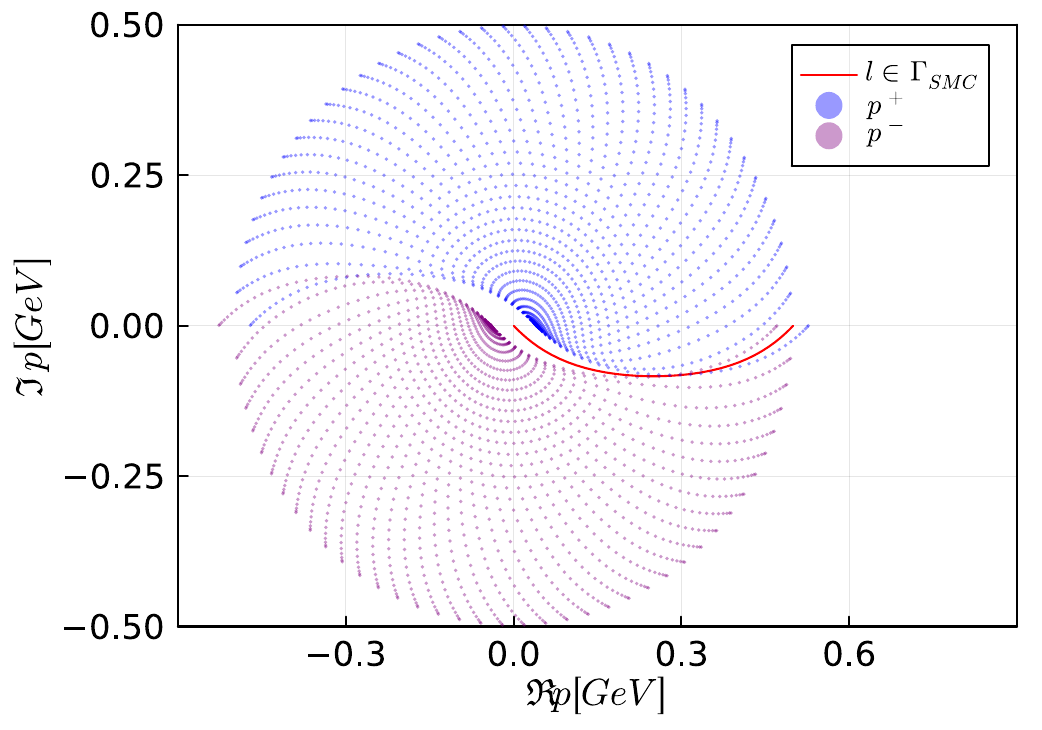}
    \caption{The complex contour $\Gamma_{\text{SMC}}$ (red line) and its singular region (gaussian points in light blue or purple) given by Eq.~(\ref{eq:singope}) for process $D^{0}D^{*+}\xrightarrow{\pi^{0}}D^{*0}D^{+}$ (left) and $D^{0}D^{*+}\xrightarrow{\pi^{-}}D^{*+}D^{0}$ (right). The parameters used in this demo are $\Lambda=0.5$ GeV, $\omega=0.1$ GeV and $V_0=-0.1$ GeV. As shown in the figures, there is no analytical problem with the left process when solving the LSE. However, crossing the cut $[-1,1]$ will occur for the right one, which can be treated properly by Eq.~(\ref{eq:acope}). See the text for further explanations.
    } 
    \label{fig:acope}
\end{figure}

In our work, we have made a proper treatment for the analytic issues as discussed above, and in Fig.\ref{fig:acope} the singular region of the integrand 
\begin{equation}
    \label{eq:singope}
    p^{\pm}=-l\cdot z\pm\sqrt{l^2(z^2-1)-m_{\pi}^2+|E_4-E_1|^2} 
\end{equation}
for $l\in\Gamma_{\text{SMC}}$ and $z\in [-1,1]$ is illustrated. We have to admit that there are still certain flaws in our numerical treatment though they do not affect what we have achieved. It is worth noting that a better treatment of the on-shell OPE is to decompose it into two parts in TOPT as in Fig.~\ref{fig:three-body2} since the OPE receives its cut only from the forward diagrams, i.e. the $DD\pi$ cuts in our work~\cite{Sadasivan:2021emk}.

\section{Correlation between cutoff parameter $\Lambda$ and potential strength $C_V$}
\label{appendix.c}
In this subsection, we give a detailed explanation of the dependence behaviour between $\Lambda$ and $C_V$ found in Fig.~\ref{fig:depcv}. To proceed, we simplify the issue of double-charm tetra-quark $T_{cc}^+$ into a single-channel scattering problem where the NLO three-body cuts are also turned off. In such a case, the pole of the amplitude, solved by a typical LSE, is evaluated by the root of the determination, namely,
    \begin{align}
        0&=1-\int_{0}^{\infty}\frac{d^3\boldsymbol{q}}{(2\pi)^3}V(\boldsymbol{q})G(\boldsymbol{q};E)\Theta(\Lambda-\boldsymbol{q})\\
        &=1+\int_{0}^{\Lambda}dq\frac{\mu_{DD^*} q^2}{\pi^2(2\mu_{DD^*} E_B+q^2)}V(q)\label{eq:singleLSE}
    \end{align}
    with $\mu_{DD^*}$ the reduced mass of the composite $DD^*$ system and $E_B\approx 273$ keV the binding energy of $T_{cc}^+$ from the LHCb's Briet-Wigner fitting as an illustration. The typical momentum scale $\gamma_T=\sqrt{2\mu_{DD^*}E_B}\approx 22$ MeV is far below a typical hard scale $0.3\sim\Lambda\sim 1.2$ GeV. The potentials involved in our work can be parametrized as,
    \begin{align}
        V(q)=c_0\cdot C_V+\sum_{i}^{N}a_i\frac{1}{q^2+\mu_i^2}+\sum_{j=1}^{M}b_j\frac{q^2}{q^2+\mu_j^2},
    \end{align}
    and then each integral term in Eq.~(\ref{eq:singleLSE}) is listed as follows,
    \begin{align}
        f^{\text{CT}}(\Lambda)&=\int_0^{\Lambda}dq\frac{\mu_{DD^*}q^2}{\pi^2(2\mu_{DD^*}E_B+q^2)}\cdot 1=\frac{\mu_{DD^*}\gamma_T}{\pi^2}[\frac{\Lambda}{\gamma_T}-\arctan(\frac{\Lambda}{\gamma_T})],\\
        f^{\text{E1}}(\Lambda,\mu_i)&=\int_0^{\Lambda}dq\frac{\mu_{DD^*}q^2}{\pi^2(2\mu_{DD^*}E_B+q^2)}\cdot \frac{1}{q^2+\mu_i^2}=\frac{\mu_{DD^*}}{\pi^2}\frac{\mu_i\cdot\arctan(\frac{\Lambda}{\mu_i})-\gamma_{T}\cdot\arctan(\frac{\Lambda}{\gamma_{T}})}{\mu_i^2-\gamma_T^2},\\
        f^{\text{M1}}(\Lambda,\mu_j)&=\int_0^{\Lambda}dq\frac{\mu_{DD^*}q^2}{\pi^2(2\mu_{DD^*}E_B+q^2)}\cdot \frac{q^2}{q^2+\mu_j^2}=\frac{\mu_{DD^*}}{\pi^2}\frac{\Lambda(\mu_j^2-\gamma_T^2)-\mu_j^3\arctan(\frac{\Lambda}{\mu_j})+\gamma_T^3\arctan(\frac{\Lambda}{\gamma_T})}{\mu_j^2-\gamma_T^2}.
    \end{align}

\begin{figure}[H]
        \centering
        \includegraphics[width=0.45\textwidth]{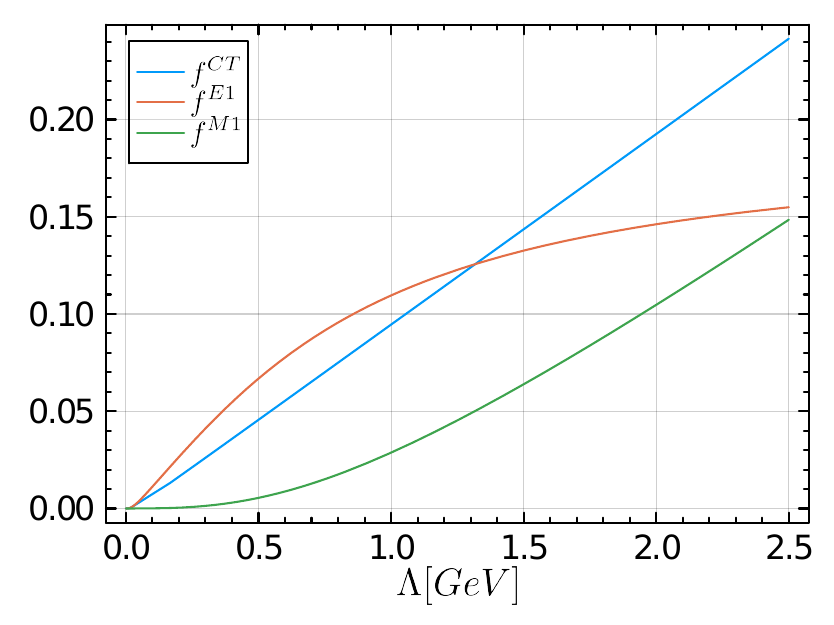}
        \includegraphics[width=0.45\textwidth]{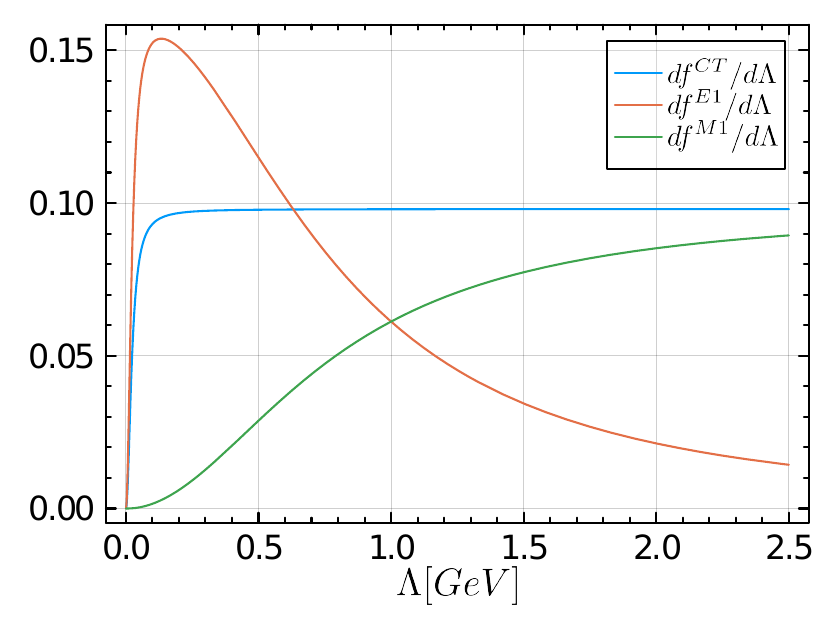}
        \caption{The functions (left) and their derivatives (right) of the three types of potentials convoluted with a two-body Green function.}
        \label{fig:fdf}
\end{figure}

Since the vector-meson-exchange plays the dominant role in the heavy meson-(anti)meson systems, we plot the above three functions and their derivatives with $\mu_{i/j}\approx m_{\rho}$ in Fig.~\ref{fig:fdf}. It shows that these three types of potentials lead to different dependence behaviors at different momentum regions: 1) that of contact potential is always proportional to $\Lambda$ due to the separation $\Lambda\gg \gamma_T$; 2) that of E1-type potentials, which manifest for a small $q$, increase with a decreasing rate; 3) that of M1-type potentials, which manifest for a large $q$, increase with an increasing rate. And then, the dependence in Fig.~\ref{fig:depcv} can be understood similarly if flipping the $\Lambda-C_V$ axes where a stationary point shows up from the competition between $f^{E1}$ and $f^{M1}$ around $\Lambda\approx 1.0$ GeV (the precise position is also related to the prefactors $a_i/b_i/c_i$).

\bibliographystyle{unsrt}
\bibliography{ref-Tcc}
\end{document}